\documentclass[12pt]{article}

\pdfoutput=1 
\usepackage{amsmath}
\usepackage{amssymb}
\usepackage{enumerate}
\usepackage[shortlabels]{enumitem}
\usepackage{graphicx}
\usepackage{hyperref}
\usepackage{xfrac}
\usepackage{float}
\usepackage{caption}
\newcommand{\unarysim}{\mathord{\sim}}
\usepackage{wrapfig}
\usepackage[section]{placeins}

\usepackage{scicite}

\title{If Loud Aliens Explain Human Earliness, \\ Quiet Aliens Are Also Rare}

\author{
\normalsize{Robin Hanson\footnote{Department of Economics, George Mason University, Fairfax, Va, USA, and Future of Humanity Institute, Oxford University, Oxford, U.K. \href{mailto:rhanson@gmu.edu}{rhanson@gmu.edu}}, Daniel Martin\footnote{Centre for Particle Theory and Department of Mathematical Sciences, Durham University, Durham, DH1 3LE, U.K. \href{mailto:contact@danielgmartin.com}{contact@danielgmartin.com}}, Calvin McCarter\footnote{Machine Learning Department, Carnegie Mellon University, Pittsburgh, PA, U.S.A, and Lightmatter Inc., Boston, MA, U.S.A. \href{mailto:calvin@lightmatter.co}{calvin@lightmatter.co}}, Jonathan Paulson\footnote{Jump Trading, U.S.A. \href{mailto:jonathanpaulson@gmail.com}{jonathanpaulson@gmail.com}}}\\
}

\date{\today}

\begin{document} 


\baselineskip14pt


\maketitle

\begin{abstract}
If life on Earth had to achieve $n$ “hard steps” to reach humanity’s level, then the chance of this event rose as time to the $n$-th power. Integrating this over habitable star formation and planet lifetime distributions predicts $>$$99$\% of advanced life appears after today, unless $n<3$ \textit{and} max planet duration $<$$50$Gyr. That is, we seem early. We offer this explanation: a deadline is set by “loud” aliens who are born according to a hard steps power law, expand at a common rate, change their volumes’ appearances, and prevent advanced life like us from appearing in their volumes. “Quiet” aliens, in contrast, are much harder to see.  We fit this three-parameter model of loud aliens to data: 1) \textit{birth power} from the number of hard steps seen in Earth history, 2) \textit{birth constant} by assuming a inform distribution over our rank among loud alien birth dates, and 3) \textit{expansion speed} from our not seeing alien volumes in our sky. We estimate that loud alien civilizations now control 40-50\% of universe volume, each will later control $\unarysim10^5$ - 3x$10^7$ galaxies, and we could meet them in $\unarysim200$Myr – $2$Gyr. If loud aliens arise from quiet ones, a depressingly low transition chance ($<\unarysim10^{-4}$) is required to expect that even one other quiet alien civilization has ever been active in our galaxy. Which seems bad news for SETI. But perhaps alien volume appearances are subtle, and their expansion speed lower, in which case we predict many long circular arcs to find in our sky.
\end{abstract}

\section{Introduction}
\label{sec:intro}

To a first approximation, there are two kinds of aliens: quiet and loud. Loud (or ``expansive'') aliens expand fast, last long, and make visible changes to their volumes. Quiet aliens fail to meet at least one of these criteria. As quiet aliens are harder to see, we are forced to accept rather uncertain estimates of their density, via methods like the Drake equation \cite{grinspoon2003natural,drake2014my, westby2020astrobiological}. Loud aliens, in contrast, are far more noticeable if they exist at any substantial density \cite{hart1975explanation}. 

Loud aliens are thus much better suited for empirical study via fitting simple models to available data. One prior researcher, S. Jay Olson, has recently pursued this approach \cite{olson2015homogeneous,olson2016visible,olson2017estimates,olson2018expanding,olson2018long,olson2020likelihood}. Olson estimated loud alien expansion speeds from the fact that we do not see now see them big in our sky, and he estimated their overall appearance rate by assuming humanity’s current date to be a representative loud alien birthdate. Olson required only one further input to estimate a full stochastic spacetime distribution over loud aliens, which was a somewhat complex appearance function, specifying the distribution of loud alien birthdates over time. 

In this paper, we explore an even simpler model of loud aliens, whom we call ``grabby,'' that we also fit to available data. Like Olson's, our model includes an expansion speed and an overall appearance rate, which we estimate similarly to Olson. But for an appearance function we use a simple power law, which has only one additional free parameter, its power $n$. This power law comes from the standard hard-steps statistical model of the origin of advanced life within a limited planetary time window, and its power comes from the number of ``hard steps'' in the evolution of advanced life, a number that has been roughly estimated from the history of major events in Earth history to be near the range 3 to 9. Yes, this simpler appearance function is a cruder approximation, but its simplicity can let us better understand and apply it.

We will show that a power law at least crudely approximates a more realistic appearance function for large fraction of the relevant parameter space. We will also show that, unless one assumes fewer than two hard steps and \textit{also} a very restrictive limit on habitable planet lifetimes, one must conclude that humanity seems to have appeared implausibly early in the history of the universe. Loud aliens, who fill the universe early, can robustly explain human earliness, as they set a deadline by which advanced life must appear if it is to see an empty universe as we do. Unlike other explanations offered, this one does not require assuming that advanced life can only result from evolutionary paths close to ours. 

As each of our three model parameters is estimable to within roughly a factor of four, we can and do estimate the stochastic spacetime distribution of grabby aliens, and thus distributions over grabby alien densities, origin dates, visible angles, and durations until we meet or see them. We also consider the alternate assumption that we have not yet noticed subtle appearance differences that mark the volumes they control, and for this scenario we estimate how common in the sky would be the volume borders for which astronomers might search.

If we assume that grabby civilizations arise from more-common non-grabby versions, then the higher the chance $p$ of this non-grabby to grabby transition, the less common can be such non-grabby versions. We will show that, assuming a generous million year average duration for non-grabby civilizations, depressingly low transition chances $p$ are needed to estimate that even one other one was ever active anywhere along our past lightcone ($p<\unarysim10^{-3}$) , has ever existed in our galaxy ($p<\unarysim10^{-4}$) , or is active now in our galaxy ($p<\unarysim10^{-7}$) . Such low chances $p$ would bode badly for humanity's future. 

\begin{figure}[h]
\centering
\begin{tabular}{l}
\includegraphics[width=4.5in]{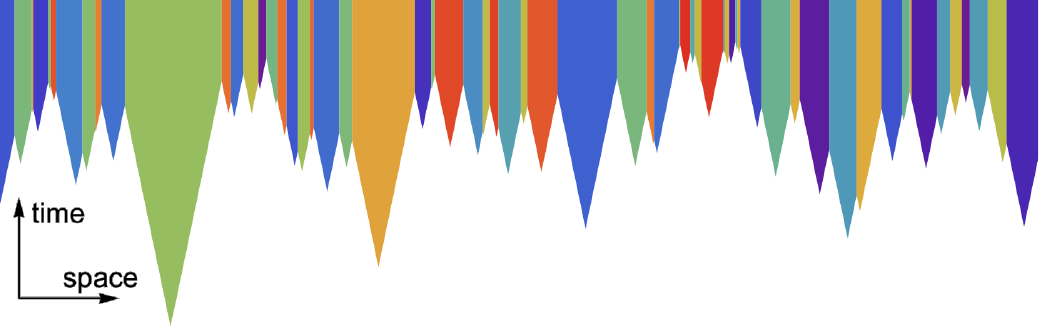} \\
\\
\includegraphics[width=4in]{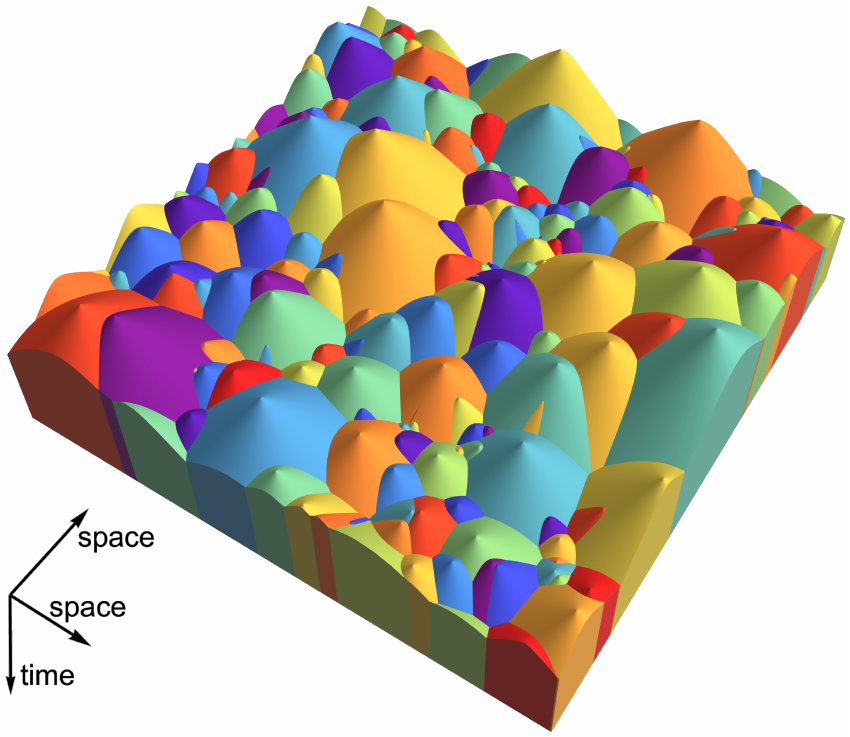} \\
\end{tabular}
\caption{These two diagrams each show a sample stochastic outcome from a grabby aliens model, in one (1D) and two (2D) spatial dimensions. The top diagram shows space on the x-axis and time moving up the y-axis. The bottom diagram shows two spatial dimensions, with time moving downward into the box. In both cases, randomly-colored grabby civilizations (GCs) are each born at a spacetime event, expand in forward-time cones, and then stop upon meeting another GC. The scenarios shown are for a clock-time volume-power $n$ of 6. This would look the same for any expansion speed, but if it is at one half lightspeed the 1D diagram shows 61 GC across a spatial width of 2.8 Gpc, while the 2D diagram shows 144 GC within a width of 0.6 Gpc, both widths applying at the median GC birthdate if that is our current date of 13.8 Gyr. We show coordinates co-moving in space and conformal in time, as explained in Section \ref{sec:cosmology}. A three-dimensional movie visualisation is found at \url{https://grabbyaliens.com}.\vspace{4pt}\\}
\label{fig:diagrams}
\end{figure}

\section{Overview}
\label{sec:overview}

Ours is a model of “grabby” aliens, who by definition a) expand the volumes they control at a common speed, b) clearly change the look of their volumes (relative to uncontrolled volumes), c) are born according to a power law in time except not within other GC volumes, and d) do not die unless displaced by other GCs. 

See Figure \ref{fig:diagrams} for examples of the space-time pattern produced by this model. GCs might allow other civilizations to be born within their volumes, but as those should clearly see that in-volume status, we exclude them from our GC definition. Our assumptions reject the ``zoo'' hypothesis \cite{ball1973zoo}.

Our model has three parameters, the first of which is the rate at which grabby civilizations are born. We assume that we humans have a non-zero chance of giving birth to a grabby civilization, and that, if this were to happen, it would happen within roughly ten million years. We also assume (with \cite{olson2016visible}) that this chance is space-time-representative, i.e., we have no good reason to expect our spacetime location to be unusual, relative to other GC origins. Given these assumptions, and the fact that we do not now seem to be within a clearly-changed alien volume, our current spacetime event becomes near a sample from the distribution of grabby civilization origins. That allows us to estimate the overall grabby birth rate to within roughly a factor of two (for its inter-quartile range), at least for powers of three or higher. 

Yes, it is possible, and perhaps even desirable, that our descendants do not become grabby. Even so, our current date remains a data point. Surprised? Imagine that you are standing on a strange planet wondering how strong is its gravity. Your intuition tells you, from the way things seem to bounce and move around you, that you could probably jump about 1.3 meters here, compared to the usual 0.5 meters on Earth. Which suggests that you are on a planet with a gravity like Mars. And it suggests this \textit{even if you do not actually jump}. A counterfactual number can be just as valid a data point as a real number. 

The second of our three model parameters is the (assumed universal) speed at which grabby civilizations expand. Our model predicts that on average at grabby origin dates, a third to a half of the universe is within grabby-controlled volumes. So if the grabby expansion speed were low, many such volumes should appear quite large and noticeable in our sky. However, as noted by (Olson ’15; Olsen ’17) and discussed in Section \ref{sec:expansionspeed}, if their expansion speed were within $\unarysim25\%$ of lightspeed, a selection effect implies that we are less likely to see than to not see such volumes. Thus if we could have seen them, they would likely be here now instead of us. As we do not now see such volumes, we conclude that grabby aliens, if they exist, expand \textit{very} fast.

The third of our three model parameters is a power derived from the effective number of ``hard steps'' in the ``great filter'' process by which simple dead matter evolves to become a grabby civilization \cite{hanson1998great}. It is well-known that the chance of this entire process completing within a time duration goes as the length of that duration raised to the power of the number of hard (i.e., take-very-long) steps (or their multi-step equivalents) in that evolutionary process. Using data on Earth history durations, a literature estimates an Earth-duration-based power to lie roughly near the range 3-9 \cite{carter1983anthropic,caldeira1992life,hanson1998early,carter2007five,watson2008implications,mccabe2011origin,aldous2012great,sandora2019multiverse,snyder2020timing,waltham2017star,lingam2019role}.

Such hard-steps power-law models are usually applied to planets. However, we show in Section \ref{sec:testpowerlaw} that such a power law can also apply well to the chances of advanced life arising within a larger volume like a galaxy. This volume-based power is our third key parameter.

For each combination of our three estimated model parameters, we can fully describe the stochastic spacetime patterns of GC activity across the universe, allowing us to estimate, for example, where they are and when we would meet or see them. We will later show in detail how these distributions change with our model parameters. 

Our grabby aliens model can explain a striking empirical puzzle: why have we humans appeared so early in the universe? Yes, many calculations often find our date to be not greatly atypical of habitable durations undisturbed by nearby sterilizing explosions, for both short and long durations \cite{ward2000rare, gonzalez2001galactic, lineweaver2004galactic, vukotic2007timescale,prantzos2008galactic,vukotic2010set,gowanlock2011model,legassick2015age,morrison2015extending,cai2021statistical,forgan2017evaluating,spinelli2021best}. Other calculations find us to be early, even if not extremely early, when planets at many lower mass stars are considered habitable \cite{lingam2018life,haqq2018we,gale2017potential,wandel2020bio}.

However, all but one of these calculations \cite{waltham2017star} neglects the hard-steps power-law. When that is included, humans look \textit{much} earlier, unless one makes strong assumptions about both the hard-steps power and the habitability of stars that last longer than our sun. (See Figure \ref{fig:rank-earth}.) We will show this via a somewhat less simple appearance model, which applies the hard-steps power law to planets, allows stars to form at different dates, and allows their planets to last for different durations. 

Our grabby aliens model resolves this puzzle by denying a key assumption of this appearance model: that the birth of some advanced life has no effect on the chances that others are born at later dates \cite{cirkovic2008astrobiological,berezin2018first}. Our model instead embodies a selection effect: if grabby aliens will soon grab all the universe volume, that sets a deadline by which others must be born, if they are not to be born within an alien-controlled volume.

Our deadline explanation for human earliness allows for a wide range of paths by which, and contexts in which, advanced life might appear. In contrast, many other explanations posit that planets around long-lifetime stars are just not habitable due to factors like tidal-locking, solar flares, insufficient early UV light, runaway greenhouses, and early loss of atmospheres and plate tectonics. However, these theories implicitly assume that advanced life can only arise through a narrow range of paths similar to Earth's path. For example, these theories fail if advanced life can arise on ocean worlds, which suffer few of these problems and comprise a large fraction of planets.

The rest of this paper will now review the robustness of the hard steps model, build a simple appearance model, use it to show how the hard-steps process makes humanity’s current date look early, describe the basic logic of our new model and how to simulate it, show how to change coordinates to account for an expanding universe, show that a power law often at least crudely approximates advanced life appearance, and finally describe our model’s specific predictions for grabby alien civilization times, distances, angles, speeds, and more, and also describe the constraints our model places on the relative frequency of non-grabby civilizations if they are to be visible to various SETI efforts. 

\section{The Hard-Steps Model}
\label{sec:hardsteps}

In 1983, Brandon Carter introduced a simple statistical model of how civilizations like ours might arise from simple dead matter, via intermediate steps of simple life, complex life, etc., a model that he and many others have since pursued \cite{carter1983anthropic,caldeira1992life,hanson1998early,carter2007five,watson2008implications,mccabe2011origin,aldous2012great,sandora2019multiverse,snyder2020timing,waltham2017star,simpson2017longevity,lingam2019role}.

Carter posited a sequence of required steps $i$, each of which has a rate $1/a_i$ per unit time of being achieved, given the achievement of its previous step. The average duration $t_i$ to achieve step $i$ is $a_i$. 

Assume that this process starts at $t=0$ when a planet first becomes habitable, and that we are interested in the unlikely scenario where all of these steps are completed by time $t=T$. (That is, assume $\sum_i t_i < T$ while $\sum_i a_i \gg T$.) Assume also for convenience that steps divide into two classes: easy steps with $a_i \ll T$, and hard steps with $a_i \gg T$.  

Conditional on this whole process completing within duration $T$, each easy step still on average takes about $a_i$, but each hard step (and also the time $T-E-\sum_i t_i$ left at the end) on average takes about $(T-E)/(n+1)$, regardless of its difficulty $a_i$. (Where $E = \sum_i a_i$ for the easy steps.) And the chance of this unlikely completion is proportional to $T^n$, where $n$ is the number of hard steps. 

It turns out that this same model was introduced in 1953 to describe the appearance of cancer, where it is now a standard model \cite{nunney2016commentary,waltham2017star}. To produce cancer, an inividual cell must host a sufficient number of mutations, and the expected time for each mutation to appear in any one cell is much longer than the organism's lifetime. Even so, cancer often appears eventually at some cell in a large organism; humans have $\unarysim 3 \times 10^{13}$ cells, and $\unarysim 40\%$ of us develop cancer.  Human cancer typically requires 6-7 hard mutations, though sometimes as few as two. The power law in timing has been confirmed, and is sometimes used to infer the number of hard mutation steps.

This basic hard steps model can be generalized in many ways. For example, in addition to these ``try-try'' steps with a constant per-time chance of success, we can add constant time delays (which in effect cut available time $T$) or add ``try-once'' steps, which succeed or fail immediately but allow no recovery from failure. These additions preserve the $T^n$ functional form. 
We can also allow ``multi'' steps where the chance of completing step $i$ within time $t_i$ goes as $t^{n_i}$ for a step specific $n_i$. If all steps have this $t^{n_i}$ form (try-once steps are $n_i = 0$, while hard try-try steps are nearly $n_i =1$), then the $n$ in the $T^n$chance rule becomes $n = \sum_i n_i$ over all such steps $i$. (And $T^n$holds here exactly; it is not an approximation.) 

For example, a ``try-menu-combo'' step might require the creation of a species with a particular body design, such as the right sort of eye, hand, leg, stomach, etc. If the length of the available menu for each part (e.g., eye) increased randomly but linearly with time, and if species were created by randomly picking from the currently available menus for each part, then the chance of completing this try-menu-combo step within time $t_i$ goes as $t_i^{n_i}$, where $n_i$ is the number of different body parts needed to be of the right sort.

We also retain this $T^n$form if we sum over different planets (or parts of planets) with different constants multiplying their $T^n$, for example due to different volumes of biological activity, or due to different metabolisms per unit volume. We could even sum over large volumes like a galaxy. Such models can allow for many sorts of “oases” wherein life might appear and evolve, not just planets. And similar models can accommodate a wide range of degrees of isolation versus mixing between these different parts and volumes.

This set of model elements can be combined in many ways to model different relevant processes. For example, a time delay plus a try-once step can roughly model a case where life on a planet dies unless it achieves a particular development within a limited time window. Such as perhaps Earth having had only a billion year window to construct habitable niches and a matching Gaian regulation system \cite{chopra2016case}. If this delay-plus-try-once process could happen in parallel with a sequence of hard steps, then this is nearly equivalent to adding a single try-once step to that sequence of hard steps.

\section{How Many Hard Steps?}
\label{sec:howmanyhardsteps}

A literature tries to estimate the number of (equivalent) hard steps passed so far in Earth’s history from key durations. Here is an illustrative calculation.

The two most plausibly diagnostic Earth durations seem to be the one remaining after now before Earth becomes uninhabitable for complex life, $\unarysim1.1$ Gyr, \cite{ozaki2021future}, and the one from when Earth was first habitable to when life first appeared, $\unarysim0.4$ Gyr (range $0.2$ to $0.8$ Gyr \cite{pearce2018constraining}). Assuming that only $e$ hard steps have happened on Earth so far (with no delays or easy steps), the expected value $a_i$ for each of these durations should be $\unarysim5.4 \textrm{Gyr}/(e+1)$. Solving for $e$ using the observed durations of $1.1$ and $0.4$ Gyr then gives $e$ values of $3.9$ and $12.5$ (range $5.7$ to $26$), suggesting a middle estimate of at least $6$.

The relevant power $n$ that applies to our grabby aliens model can differ from this number $e$. For example, it becomes smaller if evolution on Earth saw many delay steps, or big ones, such as from many easy steps, in effect reducing Earth’s $\unarysim 5.4$ Gyr time window to complete hard steps. But the relevant number of hard steps becomes larger than this $e$ if there were hard steps before Earth, or if there will be future hard steps between us today and a future grabby stage. 

Note that the enormous complexity and sophistication of even the simplest and earliest biology of which we know suggests to us that there may have been hard steps before Earth, which could be plausibly accommodated via panspermia \cite{ginsburg2018galactic,sharov2006genome,sharov2013life}. Similarly, the many obstacles to becoming grabby allow for future filters before our descendants can reach that stage. While calcuations in this paper will assume that any future filters are try-once or short delay steps, it would be straightforward (if effortful) to redo these calculations given specific assumptions about future long delays or hard steps.

Two more corrections are relevant to the power $n$ that applies to our grabby aliens model. First, Section \ref{sec:miscappendix} suggests that this paper's simplifying assumption of a power law expansion of the universe underestimates the effective volume-power regarding the space-time distributions of events by roughly a factor of three, because our actual universe has recently been transitioning toward exponential expansion. Second, Section \ref{sec:testpowerlaw} shows that while a power law also applies to volumes like galaxies, not just planets, that volume-power tends to be larger than the planet-based power.

In the following, we will take volume power $n = 6$ as our conservative middle estimate, and consider $n$ in 3 to 12 as our plausible range, though at times we also consider $n$ as low as 1 and as high as 50.

\section{Advanced Life Timing}
\label{sec:appearance}

We've just seen how power laws can describe the timing of advanced life appearance on individual planets. But how well can power laws approximate advanced life timing for much larger, perhaps galaxy-sized, co-moving volumes that contain changing mixes of planets?

To explore this question, and to estimate human earliness, we now consider a somewhat more realistic model for the timing of the appearance of advanced life. In this model, stars form at different dates, planet lifetimes vary with star lifetimes, and only planets with lifetimes $L < \bar{L}$ are suitable for advanced life. The probability density function $\alpha(t)$ of advanced life to appear at date $t$ becomes
\begin{gather}
    \alpha(t) = q \int^t_{\max(0,t-\bar{L})} (t-b)^{n-1} \varrho(b)\Big[H(\bar{L})-H(t-b)\Big] db,
    \label{eq:alpha-t-cdf}
\end{gather}
where $b$ is each star's birthdate, $\varrho(t)$ is a star formation rate (SFR), $H[L]$ is a cumulative distribution function (c.d.f.) over planet lifetimes, $n$ is a planet-based hard steps c.d.f. power, and $q$ is a normalization constant (to make $\int_0^\infty \alpha(t) = 1$).

The c.d.f. of stellar lifetimes $L$ can be approximated as roughly $H[L] = L^{0.5}$ over an important range (up to a maximum star lifetime $\bar{L} \sim 2 \times 10^4$ Gyr), because stellar mass $m$ has a c.d.f. that goes roughly as $m^{-1.5}$, while stellar lifetimes go roughly as $m^{-3.0}$ (down to $\unarysim 0.08 M_\odot$). Yes, more accurate and complex approximations exist, but this level of accuracy seems sufficient for our purposes. 

A large literature tries to estimate a general star formation rate (SFR). While this literature embraces a wide range of functional forms, a common form is $\varrho(t) = t^\lambda e^{-t/\varphi}$, which peaks at $\chi = \lambda \varphi$. And the most canonical parameter estimates in this literature seem to be power $\lambda=1$ and decay time $\varphi = 4$ Gyr \cite{madau2014cosmic,mason2017astrophysical,carnall2019measure}. However, as the SFR literature also finds a wide range of other decay times, we will below consider three decay times: $\varphi$ in $2,4,8$ Gyr, all of which we consider plausible.

For the purposes of estimating advanced life timing, we want not general SFRs $\varrho(t)$ and lifetime distributions $H[L]$, but \textit{habitable} SFRs $\varrho(t)$ and lifetime distributions $H[L]$. A habitable SFR $\varrho(t)$ selects from among all stars only those suitable for hosting advanced life. And a habitable lifetime distribution $H[L]$ describes the durations of habitability of planets, not their durations of existence.

A ``galactic habitable zone'' literature tries to estimate star habitability as a function of galactic position and time, in part by considering rates of nearby sterilizing explosions such as supernovae and gamma ray bursts \cite{ward2000rare, gonzalez2001galactic, lineweaver2004galactic, vukotic2007timescale,prantzos2008galactic,vukotic2010set,gowanlock2011model,legassick2015age,morrison2015extending,cai2021statistical,forgan2017evaluating,spinelli2021best}. 
While such explosions are much less damaging to simpler life \cite{sloan2017resilience}, in this paper we accept the usual assumption of no panspermia, and thus require habitable planets to support complex life, which is more fragile.

As this literature has focused on advising SETI efforts, it has focused mostly on dates before today. But it almost always finds peak dates $\chi$ much later than the canonical SFR peak of $\chi = 4$ Gyr, and as often as not finds peak dates after our current date of $13.8$ Gyr. We conservatively approximate typical results of this literature by retaining for our habitable $\varrho(t)$ the SFR functional form $\varrho(t) =  t^\lambda e^{-t/\varphi}$, but now with a habitable peak of $\chi = 12$ Gyr and combining this with the three plausible SFR decay times: $\varphi$ in $2, 4, 8$ Gyr. In this literature, the early rise to a peak often seems convex, fitting $\lambda > 1$. The appendix also considers varying this peak $\chi$ in $4, 8, 12$ Gyr. 

(Our approach here of representing habitability via a habitable star formation rate in effect marks most early-universe stars as entirely unsuitable for life due to overly frequent nearby explosions. A more precise calculation would consider each star's specific duration lengths between such explosions.) 

A related literature asks how the habitable planet lifetime distribution $H[L]$ relates to the raw star lifetime distribution. Many suggest that low mass star planets are not habitable due to ways in which they tend to differ from Earth. For example, such planets tend to suffer more that Earth  from tidal-locking, solar flares, insufficient early UV light, runaway greenhouses, and early loss of atmospheres and plate tectonics. Others, however, see these problems as real but not so severe \cite{lingam2018life,haqq2018we,gale2017potential,wandel2020bio}.

In Section \ref{sec:testpowerlaw}, we will show that Equation (\ref{eq:alpha-t-cdf}) is, for our purposes, often at least crudely approximated by a power law. In our grabby aliens model, we will thus assume that the chance for an advanced civilization to arise in each ``small'' (perhaps galaxy size) volume by date $t$ after the big bang goes as $t^n$. We assume that this $t^n$ form applies not just to the class of all advanced life and civilizations, but also in particular to the subclass of ``grabby'' civilizations.

\section{Are Humans Early?}
\label{sec:humansearly}

In the history of the universe, humans have arrived late relative to the peak of star formation. However, many have noted that we seem to have arrived early relative to habitable places and times. Only $\unarysim8\%$ of interstellar gas has yet been turned into stars, and if stars of all masses were equally habitable, most habitable years should lie in planets at the far-longer-lived small mass stars \cite{behroozi2015history,loeb2016relative,lingam2018life,haqq2018we,gale2017potential,wandel2020bio} After all, $95\%$ of stars last longer than our sun, and some last roughly two thousand times longer. 

Most calculations of human earliness estimate the timing of habitable planet durations, and many assume minimum duration lengths to produce advanced life. However, only one prior timing analysis \cite{waltham2017star} includes the hard steps power law that we include in Equation (\ref{eq:alpha-t-cdf}). That one prior analysis finds, as we also find here, that high powers \textit{greatly} favor later arrival dates, and so make our current date look \textit{much} earlier. (Other authors have acknowledged the relevance of the hard steps model, but not applied its power law to timing estimates \cite{lingam2019role,simpson2017longevity}.)

We now estimate human earliness as a function of the planet-based hard-steps power $n$ and planet habitability, using two model parameters to represent variation in low mass star habitability. First, we allow the maximum habitable planet lifetime $\bar{L}$ to vary from half of our sun’s lifetime of $L_\odot \approx 10$Gyr (or roughly Earth’s habitable lifetime) to the apparent max stellar lifetime of $\unarysim2 \times 10^4$ Gyr. Second, we multiply the usual star mass c.d.f. $m^{-1.5}$ by a factor $m^\kappa$, to favor larger star habitability by this factor. This changes the planet lifetime c.d.f. to go as roughly $H[L] = L^{(3-2\kappa)/6}$. (When this is unbounded, we apply a low lifetime lower bound, which turns out not to matter.) We consider 0 and 3 as values for this \textit{mass-favoring power} (MFP) $\kappa$.

\begin{figure}[h]
    \centering
    \includegraphics[width=4.5in]{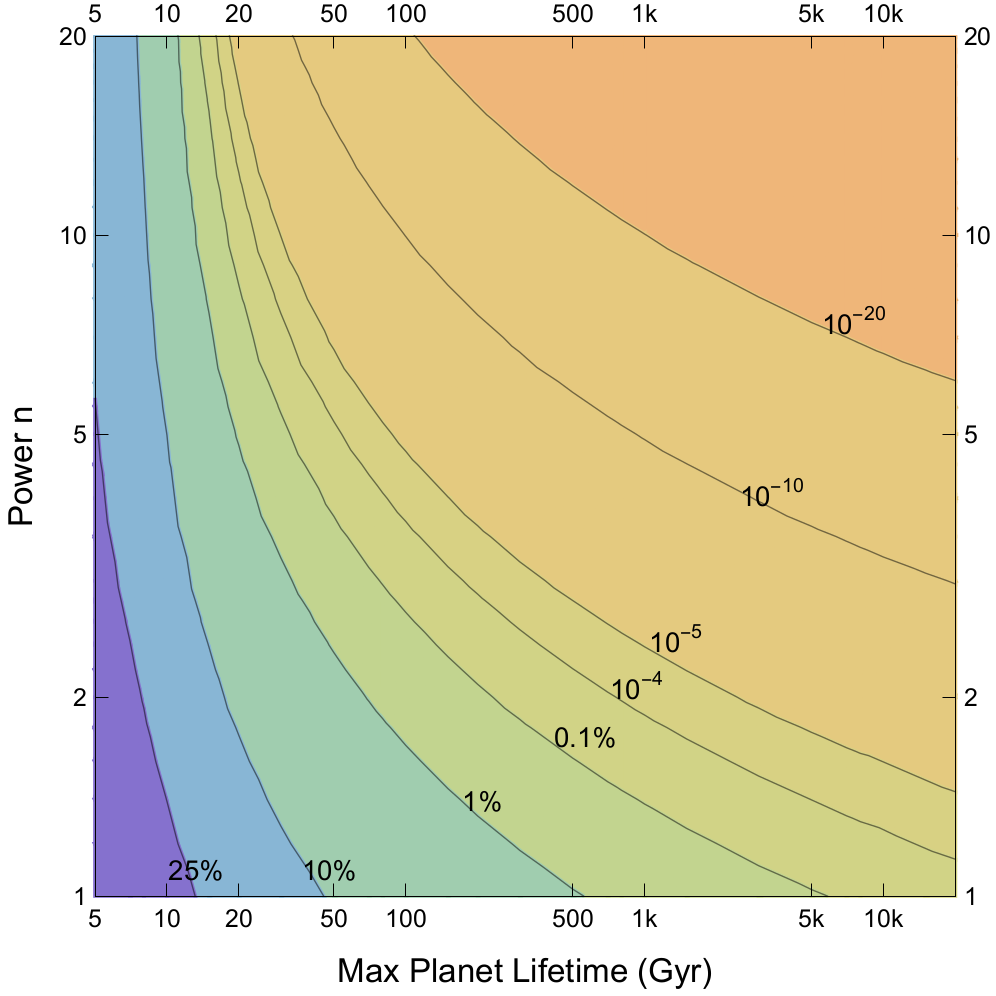}
    \caption{Percentile rank of today’s date in distribution of advanced life arrival dates, as given by Equation \ref{eq:alpha-t-cdf}, for planet-power $n$ and max habitable planet lifetime $\bar{L}$, assuming habitable SFR decay $\varphi=4$, peak $\chi = 12$ Gyr, and mass-favoring power (MFP) $\kappa=0$.}
    \label{fig:rank-earth-smooth}
\end{figure}

\begin{figure}[h]
    \centering
    \includegraphics[width=6.5in]{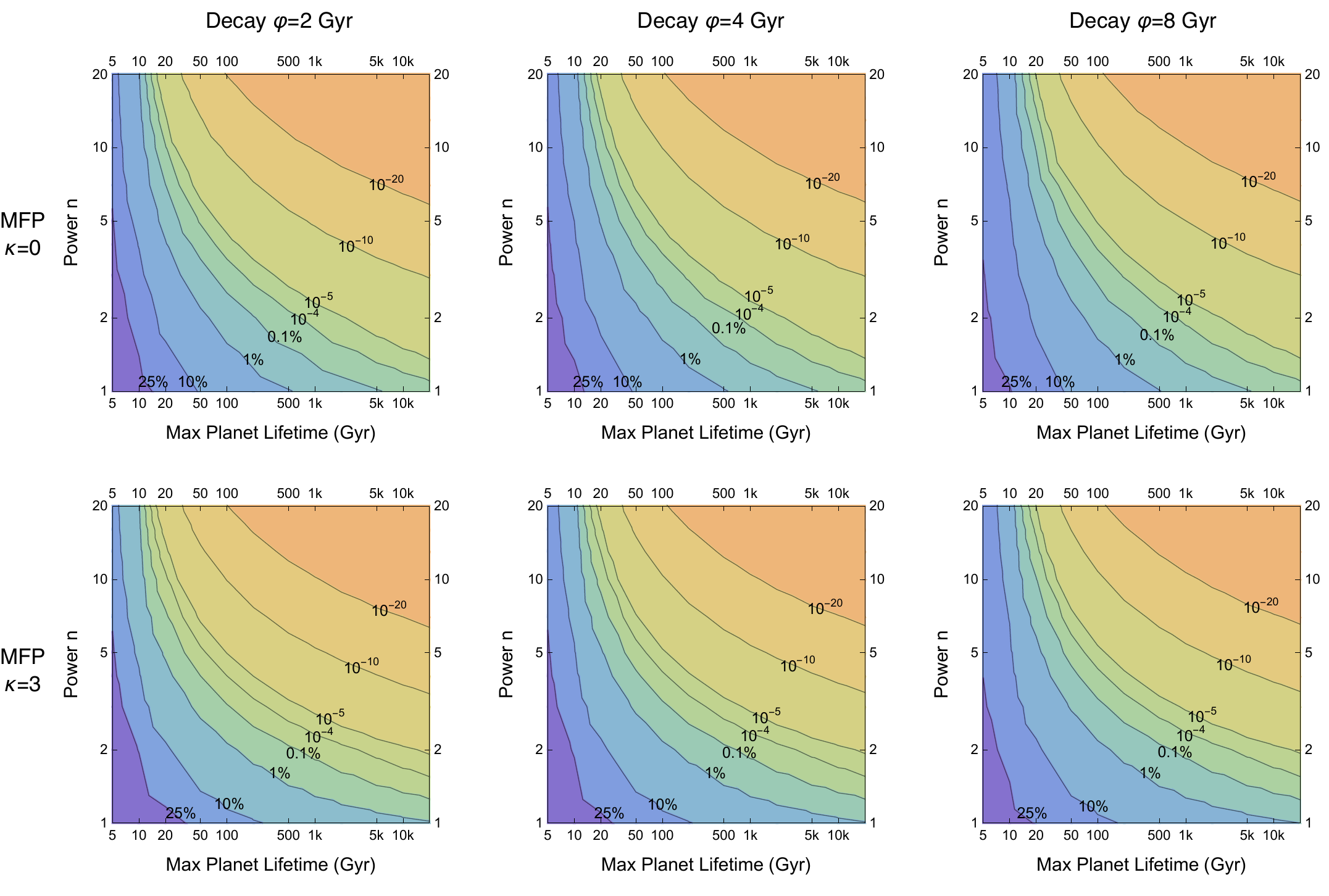}
    \caption{Percentile rank of today’s date in distribution of advanced life arrival dates, as given by Equation \ref{eq:alpha-t-cdf}. Six diagrams show six combinations of MFP $\kappa$ and SFR decay $\varphi$.}
    \label{fig:rank-earth}
\end{figure}

Figure \ref{fig:rank-earth-smooth} shows the percentile rank of today's 13.8 Gyr date within the predicted $\alpha(t)$ distribution of advanced life arrival dates, according to Equation \ref{eq:alpha-t-cdf}. It shows how this rank varies with planet-power $n$, and max planet lifetime $\bar{L}$ \cite{wolfram}. In Figure \ref{fig:rank-earth}, we also show how this percentile rank varies with SFR decay $\varphi$ and mass-favoring power $\kappa$. Section \ref{sec:miscappendix} also shows how results change as we vary the habitable SFR peak $\chi$ in $4, 8, 12$ Gyr. 

Figure \ref{fig:rank-earth-smooth} shows that for substantial powers, the effect of the power law tends to be overwhelming, while Figure \ref{fig:rank-earth} suggests that changes to habitable SFR decay $\varphi$ and MFP $\kappa$ make only modest differences relative to this strong power law effect. For MFP $\kappa=0$ our percentile rank is always below 1\% for max lifetimes $\bar{L}$ beyond a trillion years, no matter what the values of other parameters. This also holds for MFP $\kappa=3$, when planet-power $n>1.5$. 

Given our middle estimate planet-power of $n=6$, then even with a very restrictive max lifetime $\bar{L} = 10$Gyr, all percentile ranks are $<10.6\%$. For power $n=3$, at the low end of our plausible range, all ranks are $<10\%$ for max lifetime $\bar{L}=15$Gyr. And at power $n=2$ all ranks are $<8.4\%$ for max lifetime $\bar{L}= 20$ when MFP $\kappa=0$, and all are $<12.4\%$ when $\kappa=3$.

Thus we are roughly at least 10\% ``surprisingly early'' for $(n,\bar{L})$ combinations in $(6,10), (3,15), (2,20)$. And these are all \textit{very} restrictive limits on planet lifetimes. Modest increases in power $n$ or max lifetime $\bar{L}$ beyond these values quickly make our rank look \textit{quite} surprisingly early. (As Section \ref{sec:miscappendix} shows, assuming earlier SFR peaks $\chi$ allows only modestly reduces the conflict.) So unless one is willing to assume rather low powers $n$, and \textit{also} quite restrictive max planet lifetimes $\bar{L}$, there seems to be a real puzzle in need of explanation: why have we humans appeared so early in the history of the universe? 

Some have tried to explain this puzzle as due to planets at low mass stars being less habitable due to the many ways that they tend to be different from Earth. For example, such planets may tend to suffer more from tidal-locking, solar flares, insufficient early UV light, runaway greenhouses, and early loss of atmospheres and plate tectonics. But as we saw by varying MFP $\kappa$ above, merely favoring higher mass stars by large factors is not sufficient to solve the puzzle; one must instead do something closer to eliminating low mass stars entirely from consideration.  

However, to infer strict habitability limits from specific ways in which planets around low mass stars are not like Earth, such as tidal-locking, one needs to assume that advanced life can only arise through a narrow range of paths similar to the path that it took on Earth. If instead advanced life can arise through many different paths across a wide range of environments, then low-mass-star-planet conditions that make it harder for such planets to follow Earth's path do not strongly eliminate such stars from consideration. For example, a large fraction of planets seem to be ocean worlds, for which most of the mentioned problems seem largely irrelevant \cite{zeng2018planet,martin2018sea,lingam2019subsurface}. Earth seems to have long been an ocean world \cite{voosen2021ancient}. And life may also be possible around binary stars, or even near and within brown dwarves \cite{mason2015circumbinary,lingam2019brown,lingam2020prospects}.

Our loud alien hypothesis instead offers an explanation for human earliness that allows for a wide range of paths to advanced life. If loud aliens will soon fill the universe, setting a deadline by which advanced life must appear if it is to not arrive within a grabby-controlled volume, that deadline applies regardless of the type of advanced life or the path it might take to become advanced.

\begin{figure}[h]
    \centering
    \includegraphics[width=3in]{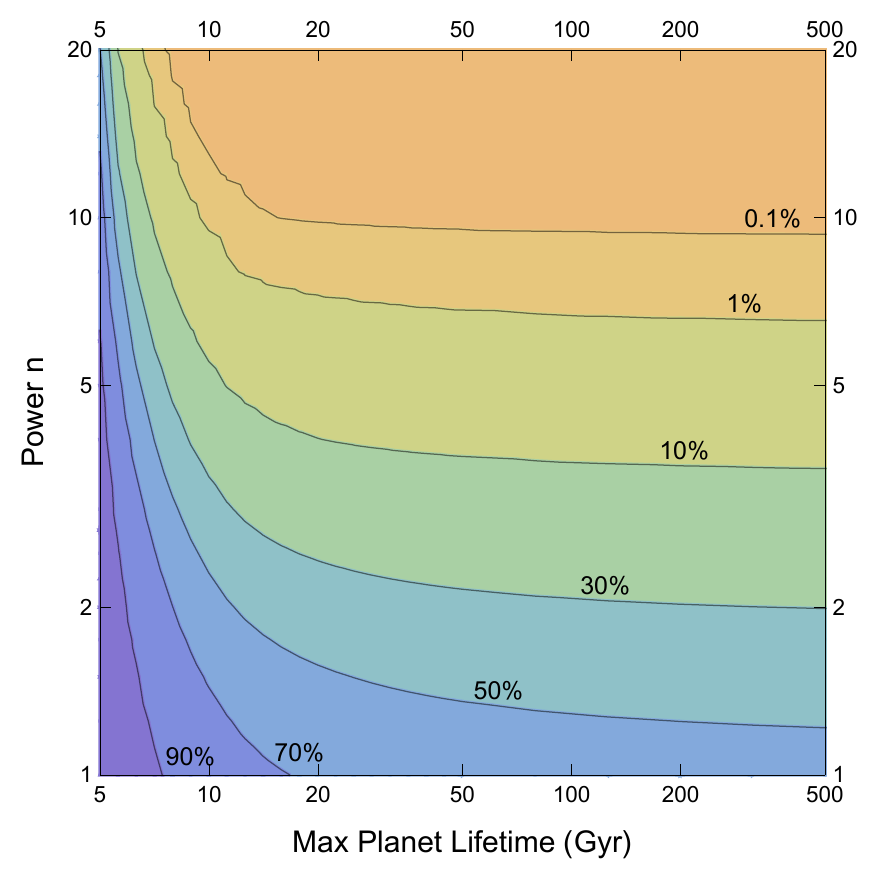}
    \caption{Fraction of advanced life star birthdates within period from Big Bang to today that are after Sun's date $b_\odot  = 9.2$Gyr, as given by Equation \ref{eq:alpha-t-cdf}, using the same parameters as in Figure \ref{fig:rank-earth-smooth}.}
    \label{fig:rank-sun}
\end{figure}

Given model parameters $n,\bar{L}$, the integral over star birthdate $b$ of Equation \ref{eq:alpha-t-cdf} can be used not only to give the percentile rank of our current date $t_0 = 13.8$Gyr, but also the percentile rank of our Sun's birthday $b_\odot  = 9.2$Gyr within the period $[0,t_0]$. Figure \ref{fig:rank-sun} shows how that rank varies with parameters $(n,\bar{L})$ for our standard values of other parameters. For maximum lifetimes $\bar{L}$ over $\unarysim20$ Gyr, less than $\unarysim10\%$ of stars are born as late as our Sun, unless power $n \leq \unarysim4$. This argues  for either relatively low planet-powers $n$, or for panspermia, which favors higher powers and requires a later Sun birthdate to fit in a earlier star with a planet at which some hard steps were achieved before Earth.

\section{Model Rationale}
\label{sec:rationale}

While our grabby alien model should ultimately stand or fall on how well it accounts for observations, readers may want to hear plausibility arguments regarding its key assumptions. 

We have already discussed (in Section \ref{sec:hardsteps}) reasons to expect a power law time dependence in the chances to originate advanced life in any one place, due to evolution needing to pass through several difficult steps. We will also suggest (in Section \ref{sec:testpowerlaw}) that this kind of dependence often applies sufficiently well not just to planets, but also to large volumes like galaxies. But why might there exist civilizations who expand fast, steadily, and indefinitely, changing how their volumes look in the process?

In Earth history, competing species, cultures, and organizations have shown consistent tendencies, when possible, to expand into new territories and niches previously unoccupied by such units. When such new territories offer supporting resources that can aid reproduction, then behaviors that encourage and enable such colonization have often been selected for over repeated episodes of expansion \cite{hudson2016lonely}. (Note that individual motives are mostly irrelevant when considering such selection effects.)

In addition, expansions that harness resources tend to cause substantial changes to local processes, which induce changed appearances, at least to observers who can sufficiently see key resources and processes. While these two tendencies are hardly iron laws of nature, they seem common enough to suggest that we consider stochastic models which embody them. 

Furthermore, when uncoordinated local stochastic processes are aggregated to large enough scales, they often result in relatively steady and consistent trends, trends whose average rates are set by more fundamental constraints. Examples include the spread of species and peoples into territories, diseases into populations, and innovations into communities of practice. Without wide coordination, processes that halt or reverse their spread locally only rarely stop their spread across very wide scales. 

Yes, expanding into the universe seems to us today a very difficult technical and social challenge, far beyond current abilities. Even so, many foresee a non-trivial chance that some of our distant descendants may be up to the challenge within ten million years. Furthermore, the large distances and times involved suggest that large scale coordination will be difficult, making it more plausible that uncoordinated local processes may aggregate into consistent overall trends. The spatial uniformity of the universe on large scales, and competitive pressures to expand faster, also suggest that such trends could result in a steady and universal expansion speed.

Yes, perhaps there is only a tiny chance that any one civilization will fall into such a scenario wherein internal selection successfully promotes sustained rapid overall expansion. (We discuss such chances in Section \ref{sec:SETI}.) Even so, the few exceptions could have a vastly disproportionate impact on the universe. If such expansions are at all possible, we should consider their consequences.

\section{The Model}
\label{sec:model}

Our basic model sits in a cosmology that is static relative to its coordinates. (An expanding cosmology is addressed in Section \ref{sec:cosmology}.) That is, galaxies sit at constant spatial position vectors $v$ in a $D$-dimensional space, time moves forward after $t=0$, and movement at constant speed $s$ in the $x$ coordinate direction satisfies $s = \Delta x / {\Delta t}$. (We are mainly interested in $D=3$, but will at times consider $D$ in $1, 2$.) 

Within this space, ``grabby civilizations'' (GC) spontaneously arise at spacetime events $(v,t)$. By definition, GC volumes look clearly different, and consistently expand in every direction at a constant local speed $s$ until meeting volumes controlled by other GC. 

Grabby civilizations are presumed to arise from non-grabby civilizations (NGC) within a short duration (less than ten million years) of the NGC's birth \cite{olson2018expanding}. We assume that NGCs which do not give birth to GCs have little cosmic impact, and thus do not block GCs from the activities that define them: being born, expanding, and changing volume appearances. Though we at times presume that humans today count as a NGC, we otherwise purposely say little about how exactly to define NGCs, so that many possible definitions can be applied. 

We assume that only one in $R$ such NGCs gives birth to a GC, and that this rate $1/R$ is independent of the other parameters in our model, at least for NGCs who are not born within a GC-controlled volume. We will later show how this ratio $R$ relates to the chances that each NGC could see evidence of other NGCs in various spacetime regions.

We assume that once a volume is controlled by any GC, it is forever controlled by some GC, and forever looks different from non-GC-controlled volumes. We make no further assumptions about what happens after GCs meet, as our analysis will only consider GC and NGC origin events where residents can see that they are not in a GC-controlled volume. Once they meet, GCs might fight over volumes, or maintain peaceful borders at their meeting locations. Each GC might even, soon after birth, trigger a false vaccum decay, in effect destroying the universe, though this scenario requires expansion at lightspeed (credit for this suggestion to Adam Brown).

We assume that each ``small'' (perhaps galaxy-region-sized) volume has the same uniform per-volume chance of a GC being born there, a chance that is independent of the chances in other volumes \cite{olson2015homogeneous}. As this uniformity ignores the actual spatial clustering of galaxies, which is mainly on scales below $30$ Mpc, we expect our model to be less accurate on such smaller scales.

We assume that over time the c.d.f. chance of birth by $t$ at some position $v$ rises as $(t/k)^n$, a volume-power $n$ of time since $t=0$ divided by a timescale constant $k$. Except that this per-time chance falls to zero as soon as the expanding volume of another GC includes this position $v$. There may in fact be civilizations born within GC volumes, but we decline to call them ``grabby'' to keep the GC concept relevant to human observations. As we discuss in Section \ref{sec:testpowerlaw}, this power law dependence can be a robust feature of the origin of advanced civilizations. 

And that is our whole model. It has three free parameters: the speed $s$ of expansion and the constant $k$ and volume-power $n$ of the appearance power law. It turns out that we can estimate each of these parameters at least roughly from data.

Specific examples of the spacetime distribution resulting from this process are shown in Figure \ref{fig:diagrams}. Notice how smaller GC tend to be found at later origin times in the spatial ``crevices'' near where larger earlier GC would meet. (This correlation between origin date and size is explored in Figure \ref{fig:origin-volume}.) For simplicity, these examples show each GC retaining control of its initial volume after GCs meet. This assumption is made for concreteness only; our analysis only depends on it when we show examples or find distributions over final GC volumes.

\section{Heuristic 1D Model}
\label{sec:heuristicmodel}

A simple deterministic model gives a rough approximation to this stochastic model in one dimension.

\begin{figure} 
\centering
\includegraphics[width=4in]{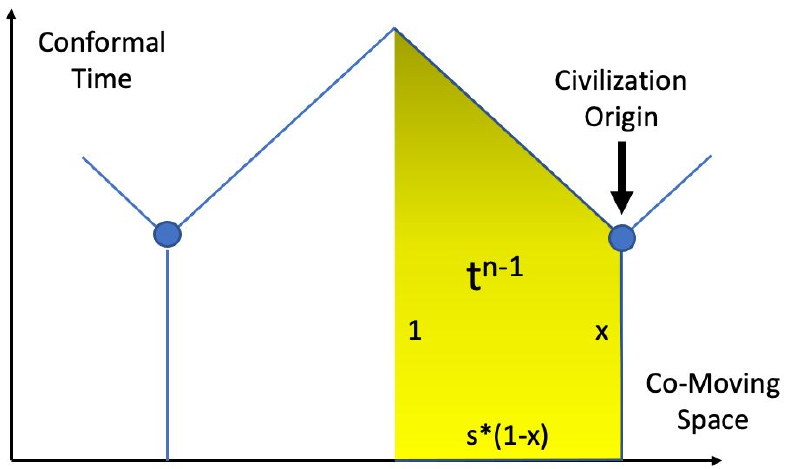}
\caption{Illustration of the heuristic math model of Section \ref{sec:heuristicmodel}}.
\label{fig:illustration-heuristic}
\end{figure}

Assume that a regular array of ``constraining'' GC origins all have the same origin time $t=x$, and are equally spaced so that neighboring expansion cones all intersect at $t=1$. (See Figure \ref{fig:illustration-heuristic}.)  If these cones set the deadline for the origins of other ``arriving'' GC, we can then find a distribution over arriving GC origin times that results from integrating the arrival power law $t^{n-1}$ over the regions allowed by the constraining GC.

The key modeling assumption of this simplified heuristic model is to equate the constraining GC origin time $x$ with the \textit{percentile rank} $r$ of the resulting distribution of arriving GC origins. This assumption implies
\begin{gather*}
\frac{1-r}{r}\int^x_0 t^{n-1}dt = \int^1_x t^{n-1}\frac{1-t}{1-x}dt
\label{eq:heuristic}
\end{gather*}
which is independent of speed $s$ or constant $k$.  This math model thus captures two key symmetries of our stochastic model, which are described in Section \ref{sec:simmodel}.

For each volume-power $n$, there is some matching rank $r$ where this heuristic model’s prediction of $(1-x)/x$ for the ratio of median time till meet aliens to the median GC origin time equals a 1D simulation result for this ratio. This matching rank is $\unarysim0.88$ at $n = 1$, falls to a minimum of $\unarysim0.61$ at $n=4$, and then rises up to $\unarysim0.88$ again at $n=24$. This simple heuristic math model thus roughly captures some key features of our full stochastic model, such as having an overall stochastic shape that is independent of speed $s$ and constant $k$.

\section{Cosmology}
\label{sec:cosmology}

Our model seems to have a big problem: its cosmology has key objects maintaining constant spatial relations, yet our universe is expanding. Our rather standard solution: a change of coordinates.

Usually, using ordinary local ``ruler'' spatial distances $dv$ and ``clock'' time differences $dt$, the spacetime metric distance $d$ between events is (for lightspeed $c=1$) given by $d^2 = dt^2 - dv^2$. We instead use ``model'' coordinates, which are co-moving spatial positions $du = dv/a(t)$ and conformal times $d\tau = dt/a(t)$. Here $a(t)$ is a ``scale factor'' saying how much the universe has expanded at time $t$ relative to time $t=1$. Metric distance then becomes $d^2 = a^2(t)\times (d\tau^2 - du^2)$. 

(Note that this coordinate change preserves local speeds and also relative angles between spatial positions. We can thus calculate such things in whichever coordinate system is more convenient.) 

In terms of our model spatial coordinates $u$, galaxies tend to stay near the same spatial positions. However, in an expanding universe a freely-falling object that starts at an initial speed $\Delta u / \Delta\tau$, and has no forces acting on it, does \textit{not} maintain that $\Delta u / \Delta\tau$ coordinate speed as the universe expands. It instead slows down \cite{carroll2004gr}. Does this show that a GC which might in a static universe expand at a constant speed $\Delta u / \Delta\tau$ does not in fact expand at such a constant speed in an expanding universe?

No, because the frontier of an expanding civilization is less like an object thrown and more like the speed of a plane; a given scenario of wing drag and plane engine power will set a plane speed relative to the local air, not relative to the ground or its initial launch. Similarly, a civilization expands by stopping at local resources, developing those resources for a time, and then using them to travel another spatial distance \cite{hanson1998burning}. As this process is relative to local co-moving materials, it does maintain a constant model speed $\Delta u / \Delta\tau$. 

We run our simulations in a static model space $u$, and in model time $\tau$. To convert our results from model time $\tau$ to clock time $t$, it suffices to know the scale factor function $a(t)$. This scale factor $a(t)$ went as $t^{1/2}$ during the ``radiation-dominated'' era from the first second until about 50{,}000 years after the big bang, after which it went as $t^{2/3}$ during the ``matter-dominated'' era. In the last few billion years, it has been slowly approaching $e^{\Omega t}$ as dark energy comes to dominate. 

Assume that the scale factor is a power law $a(t) = t^m$, and that we today are at percentile rank $r$ in the distribution over GC origin dates. If so, we can convert from model time $\tau$ to clock time $t$ via $t = t_0(\tau/{\tau_0})^{1/(1-m)}$ and $r=F(\tau_0)$, where $t_0 = 13.787$ Gyr is a best estimate of the current age of the universe, and $F(\tau)$ is the c.d.f. over GC model origin times. We can also convert between clock-time volume-power $n$ and model-time volume-power $\eta$ via $\eta = n/(1-m)$.

Now, a least-squared-error fit of a power law to the actual $a(t)$ within 0-20 Gyr after the Big Bang gives a best fit of $m \approx 0.90$. However, as this $m$ value tends to give more extreme results, we conservatively use $m = 2/3$ in most of our analysis. In Section \ref{sec:miscappendix}, we show how some results change for $m = 0.9$.

Note that by assuming a uniform distribution over our origin rank $r$ (i.e., that we are equally likely to be any percentile rank in the GC origin time distribution), we can convert distributions over model times $\tau$ (e.g., an $F(\tau)$ over GC model origin times) into distributions over clock times $t$. This in effect uses our current date of 13.8Gyr to estimate a distribution over the model timescale constant $k$. If instead of the distribution $F(\tau)$ we use the distribution $F_0(\tau)$, which considers only those GCs who do not see any aliens at their origin date, we can also apply the information that we humans do not now see aliens. 

In the rest of this paper we will show spacetime diagrams in terms of model coordinates $(u,\tau)$, but when possible we will discuss and show statistics and distributions in terms of clock times $t$, including clock-time volume-powers $n$. To describe distributions of events in space within an expanding universe, we will focus when possible on counting galaxies, as their number seems to be largely conserved on our timescales of concern, and galaxies tend to stay near the same co-moving spatial coordinates $u$.  

At the current date, our universe now has $g = \unarysim7 \times 10^7$ galaxies per Gpc$^3$ \cite{conselice2016evolution}. (This applies to galaxies defined as star clusters with mass $> 10^6 M_\odot$; there are $~\unarysim7$ times more galaxies if we define them as mass $> 10^5 M_\odot$.) As the average stars per galaxy is$~\unarysim 10^8$, while our galaxy has $~\unarysim 10^{11}$, our Milky Way galaxy should count as $M = \unarysim 10^3$ average galaxies for events we expect to happen per star, such as GC and NGC births. 

If we identify our current date of $t_0 = ~13.787$ Gyr with model time $\tau$ in $[0,1]$, then a speed $s=1$ model box $[0,1]^3$ corresponds to proper volume of $(t_0 /\tau)^3(s/c)^3 \textrm{Gpc}^3$ at that time $t_0$. Thus the model box holds $G = g (t_0 /\tau)^3(s/c)^3$ galaxies at \textit{all} model and clock times. And the model volume corresponding to one average galaxy is at all times $1/G$.

\section{Simulating The Model}
\label{sec:simmodel}

Our stochastic model scales in two ways. That is, two kinds of transformations preserve its stochastic pattern of GC space-time origins, when expressed in model coordinates. First, halving the speed $s$ of expansion halves the average spatial distance between GC, but otherwise preserves their relative spacetime pattern. (Though this can change parameters that depend on this speed $s$ relative to $c$, such which GC can see what other GC.) Second, changing timescale $k$ changes the median GC origin time, but preserves the pattern once times and distances are rescaled by the same factor to give the same median origin time as before. (This even preserves who-can-see-who relations.)

Thus simulations need only vary dimension $D$ and volume-power $n$, and repeatedly sample, to see the full range of stochastic GC origin patterns that can be produced by this model. This ability to explore its full range is one of the main virtues of our admitedly oversimplified model.

So we can fix expansion speed at $s=1$, focus on a unit time range $[0,1]$ and spatial volume $[0,W]^D$, and use a ``wrap around'' (toroidal) metric which identifies $x=0$ with $x=W$, etc. in all spatial dimensions. ($W \geq 1$ seems wise, and $W=1$ usually seems fine; in Section \ref{sec:miscappendix} we check for robustness to this choice.) We generate $N$ candidate GC origins $(u,\tau)$ as (uniformly) random positions $u$ within this unit volume, paired with random times $\tau$ drawn from a c.d.f that goes as $\tau^n$ on $[0,1]$. 

Let us say that, for $s=1$, spacetime event $A$ ``precludes'' event $B$ if $A$’s time is earlier and if the spatial distance between them is less than their time difference. Given a set of $N$ candidate origin events, we filter out any members precluded by other members, and the remaining set $C$ of origins then defines a stochastic sample from our model. (It helps to test earlier candidates first, each against the test-passing origins collected so far.) 

We then rescale all times and distances in $C$ by the same factor, to make the median origin time be one. We can then transform such a sample into a sample with a differing speed $s$ by rescaling all distances by the same factor. And we can transform it into samples with differing timescales $k$ by rescaling all times and distances by the same factor. (Such ``samples'' may describe the same basic stochastic pattern over larger or smaller spatial volumes, in essence holding more or less ``copies'' of the same basic stochastic pattern.)

We know that a spacetime event is controlled by some GC if it is precluded by any GC origin. While a larger sample of $N$candidate GC origins tends to induce a larger non-precluded set $C$, eventually $C$ stops increasing substantially with $N$, giving a ``full'' sample. 

Lightspeed $c$ can be varied relative to speed $s$ to calculate who can see who in such a sample. When $c$ is large compared to $s$, calculations of what can be seen must look outside the model box to consider an indefinite array of exact copies of that box and its contents in every direction.

\section{Simulation Statistics}
\label{sec:simstats}

Assume one has a sample of simulation runs, each of which tried $N$ candidates and produced a set of $C$ grabby civilizations at origins $(u,\tau)$ within the model box $[0,W]^3$.  Here are some interesting statistics that one can calculate within each run (and average over multiple runs). 

The following statistics describe how GCs are distributed in space and time: 

\begin{enumerate}[A)]
\item We can collect a distribution $F(\tau)$ over model GC \textit{origin times} $\tau$. (This is independent of speed $s$.) 
\item Given a set of GC model origin times, if we assume Earth today has its GC origin rank $r$, that gives a constant for converting all model times in that simulation into clock times. If we then assume a uniform distribution over Earth rank $r$ (expressing the assumption that we have a spacetime-representative chance of birthing grabby descendants soon), we can convert any distribution over model times into a distribution over clock times. We can also take any distribution over pairs of model times into a distribution over clock durations between those times. (This is independent of speed $s$.) 
\item If at some date, the model volume wherein a particular GC is first to arrive is $V < 1$, then it covers $VG$ galaxies at that date, at least if we assume that GCs who meet simply stop and keep control of their prior volumes \cite{olson2018long}. (Alternatively, we might call this a GC's ``volume when meet neighbors''.) Iterating through the GC, a distribution over galaxies per GC can be found for different times, as can the \textit{volume fraction} of the universe which is controlled by GCs at that GC origin time $\tau$. (This fraction is independent of speed s.) 
\item We can collect a two-dimensional distribution showing the relation between GC origin times and galaxies per GC.
\end{enumerate}

The following statistics describe what we may see in our future regarding GCs: 

\begin{enumerate}[A)]
\addtocounter{enumi}{4}
\item For each GC origin position, we can find the model time $\tau$ at which a speed $s=1$ traveler would arrive at that position from each other GC origin event. The minimum of these is the \textit{arrival time} when the first other GC expansion wave would, if allowed, arrive at this GC position. (This is independent of speed $s$.) 
\item The model time average of that min arrival time and this GC origin time is the \textit{meet time}, when the two GC expansion waves would collide (if both were in fact GCs). The first GC it meets is also the first one to arrive. (This is independent of speed $s$.) 
\item If a GC has not yet seen any other GC, it will first see the GC that it will first meet, and see it before their meeting for $s < c$. For $s=1$ that \textit{when see time} is $\tau = (d + \tau_0 + c \tau_1)/(1+c)$, where $\tau_0$ is the viewer’s origin time, and $\tau_1$ is the viewed’s origin time.
\item We can convert origin and view times from model times into clock times, take the difference and get a distribution over clock \textit{time until see} aliens. (That duration is zero if aliens can already be seen from a GC origin.)
\end{enumerate}

The following statistics describe what we may see now regarding GCs: 

\begin{enumerate}[A)]
\addtocounter{enumi}{8}
\item If $b$ and $d$ are the (model) time and space distances between two GC origins, then for $s=1$ if $d<cb$ the earlier one would be visible to the later one. The earlier GC-controlled-volume would (unless it had already collided with another GC) appear as a disk in the sky to the later one, with angle $\theta$ given by $\tan(\theta/2) = x/(c(b-x))$, where $x$ solves $d^2 = x^2 + (c(b-x))^2$. We can thus find a distribution over the \textit{max angle} that each GC can see \cite{olson2016visible}. (If the GC sees none, its max angle is zero.)
\item We can find a distribution over GCs of the number of other GC origins that each GC can see at its origin. (This assumes transparent GC volumes.)
\item A GC that appears as a disk of angle $\theta$ in the sky will have a border $B(\theta)$ of angular length in the sky $B(\theta) = 2^{3/2}\pi(1-\cos(\theta/2))^{1/2}$. If GCs are transparent we can add up these border lengths over visible GCs to find a total border length visible at each GC origin, and then collect that into a distribution over all GC.
\item We can select the subset of GCs who don’t see any other GC-controlled-volume at their origin. This GC set can be used to compute other statistics conditional on not seeing GCs.
\end{enumerate}

The following statistics describe what we may see now regarding NGCs: 

\begin{enumerate}[A)]
\addtocounter{enumi}{11}
\item An NGC to GC ratio $R = (g t_0^3/MN)(s/c)^3 {\tau^{-\eta-3}}$ is required to expect one prior NGC in the same Milky-Way-sized galaxy in the past light cone of a GC (or NGC) origin there at (model) time $\tau$. Iterating through GC origin dates $\tau$ gives a distribution of such ratios.
\item To be visible from Earth today, an NGC in our galaxy with a lifetime of $L$ would have been born in clock time window of width $[t_0-L,t_0]$, which for $L \ll t_0$ is model time range $[\tau(1-(L/(3t_0)),\tau]$ , at least when $m = 2/3$. An NGC to GC ratio $R = (3g t_0^2/MNL(\eta-1))(s/c)^3 {\tau^{-\eta-2}}$ is required to expect an active NGC visible in a Milky-Way-sized galaxy at a GC (or NGC) origin there at $\tau$.
\item An NGC to GC ratio $R = (1/8\pi N)(\Gamma(\eta+4)/(\Gamma(\eta)(\eta-1)))(s/c)^3 {\tau^{-\eta-3}}$ is required to expect one prior NGC anywhere in the past lightcone of a GC (or NGC) origin at time $\tau$. 
\item An NGC to GC ratio $R = (1/8\pi N)\eta(\eta-2)(s/c)^2 (3t_0/L){\tau^{-\eta-3}}$ is required to expect one NGC whose active period intersects with the past lightcone of a GC (or NGC) origin at time $\tau$. 
\end{enumerate}

Code to simulate the grabby alien model and compute the above statistics can be found at \url{https://github.com/jonathanpaulson/grabby_aliens}.

\section{Estimating Expansion Speed}
\label{sec:expansionspeed}

\begin{figure} 
    \centering
    \includegraphics[width=4in]{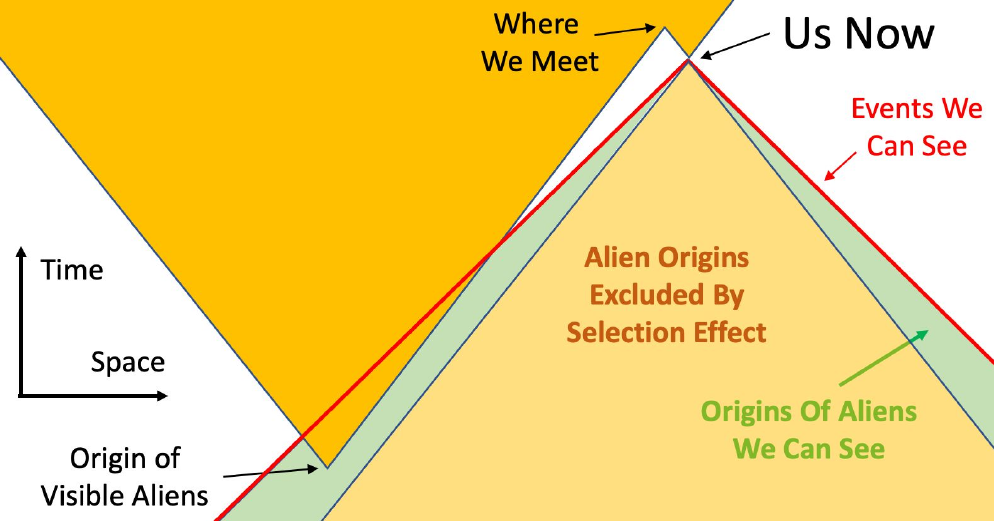}
    \caption{Illustration of selection effect. When expansion speeds $s$ are near lightspeed $c$, GC origin events in most of our past lightcone would have created a GC that controls the event from which we are viewing, preventing us-now from becoming a candidate GC origin \cite{olson2017estimates}} 
    \label{fig:illustration-selection}
\end{figure}

Our grabby aliens model has three free parameters, and we have so far discussed empirical estimates of two of them: the volume-power $n$ (in Sections \ref{sec:howmanyhardsteps} and \ref{sec:testpowerlaw}) and the power law constant $k$ (in Section \ref{sec:cosmology}). The remaining parameter, expansion speed $s$, can also be estimated empirically, via the datum that we humans today do not see alien volumes in our sky \cite{olson2015homogeneous,olson2016visible,olson2017estimates}. 

\begin{figure}
\centering
\begin{minipage}{.45\textwidth}
\centering
\includegraphics[width=1.0\linewidth]{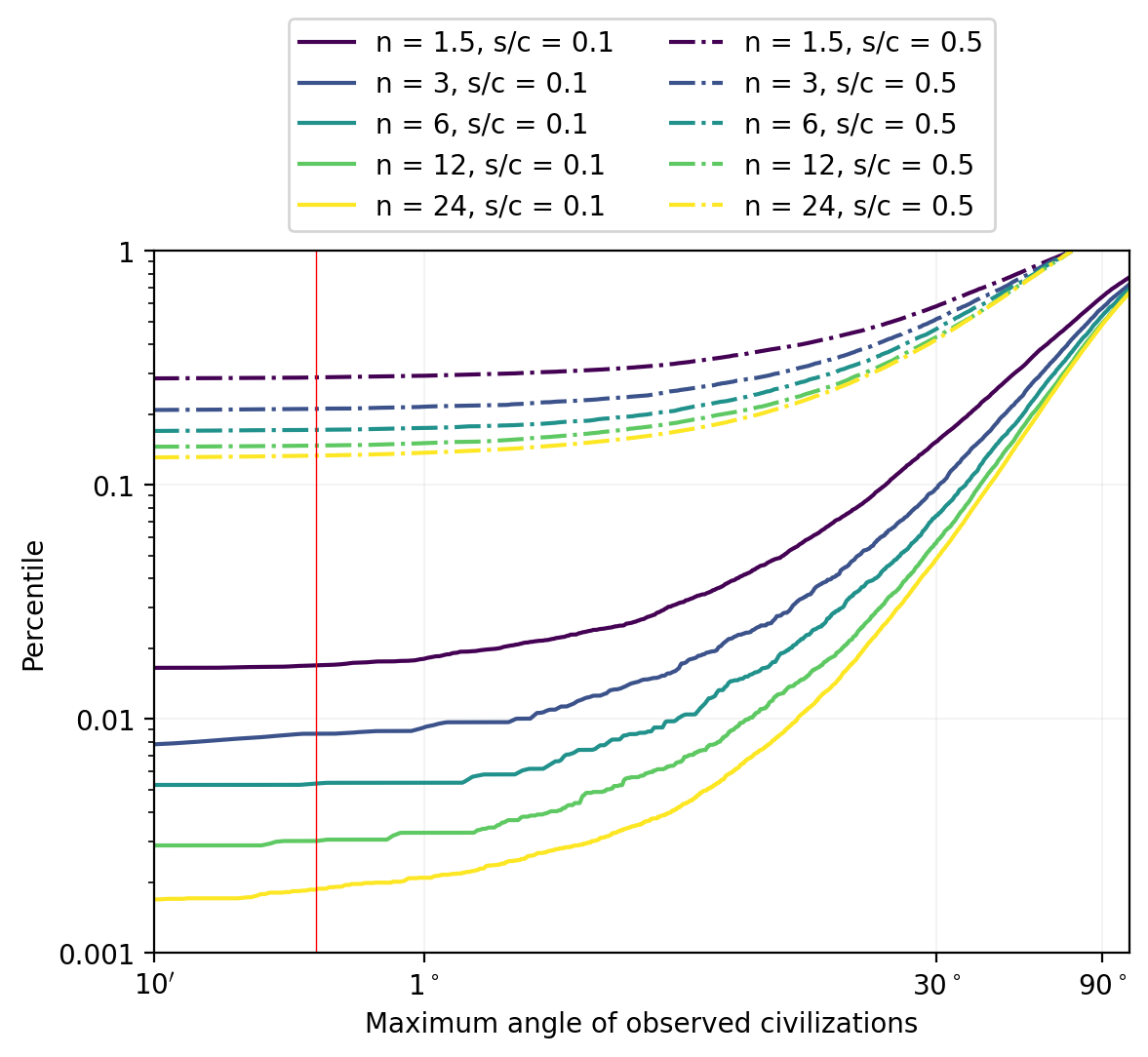}
\caption{C.d.f.s over largest angle in sky of GC seen from GC origins. Red line shows our Moon’s diameter $(29'20'')$.}
\label{fig:cdf-max-angle}
\end{minipage}\qquad\quad\begin{minipage}{.45\textwidth}
\vspace{10pt}\includegraphics[width=1.05\linewidth]{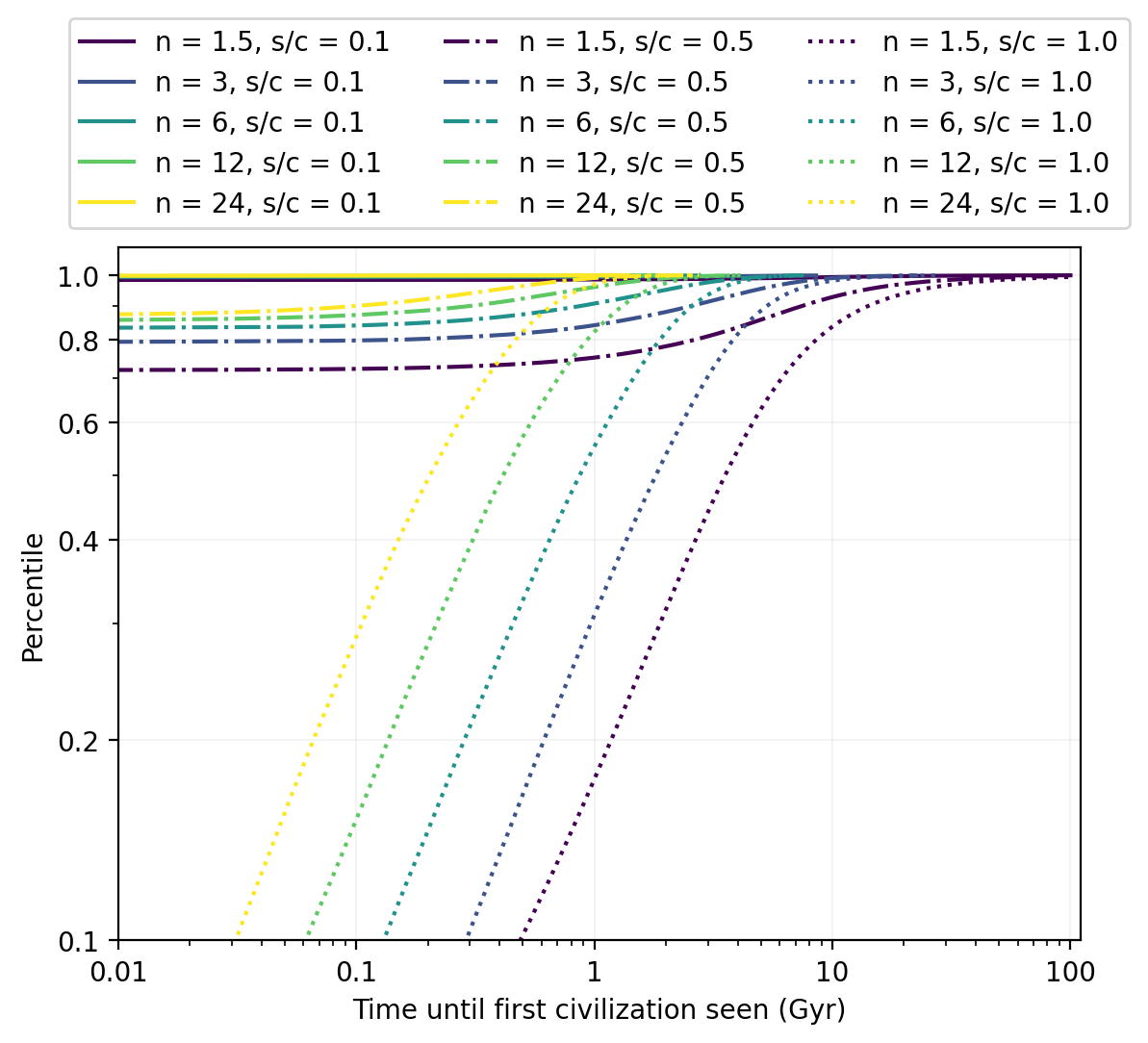}
\caption{C.d.f.s over clock time till some GC descendant sees aliens. For $s = 0.1c$, almost all GC have already seen aliens.}
\label{fig:cdf-time-see}
\end{minipage}
\end{figure}

We will see in Figure \ref{fig:cdf-max-angle} that visible alien volumes are typically huge in the sky, much larger than the full moon. So there is only a tiny chance of a visible GC volume being too small to be seen by the naked eye, much less by our powerful telescopes. So if alien volumes looked noticeably different, as we assume in our definition of ``grabby'', and if their volumes intersected with our backwards light cone, then we should see them clearly. 

\begin{figure} 
\centering
\begin{minipage}{.45\textwidth}
\centering
\includegraphics[width=1.05\linewidth]{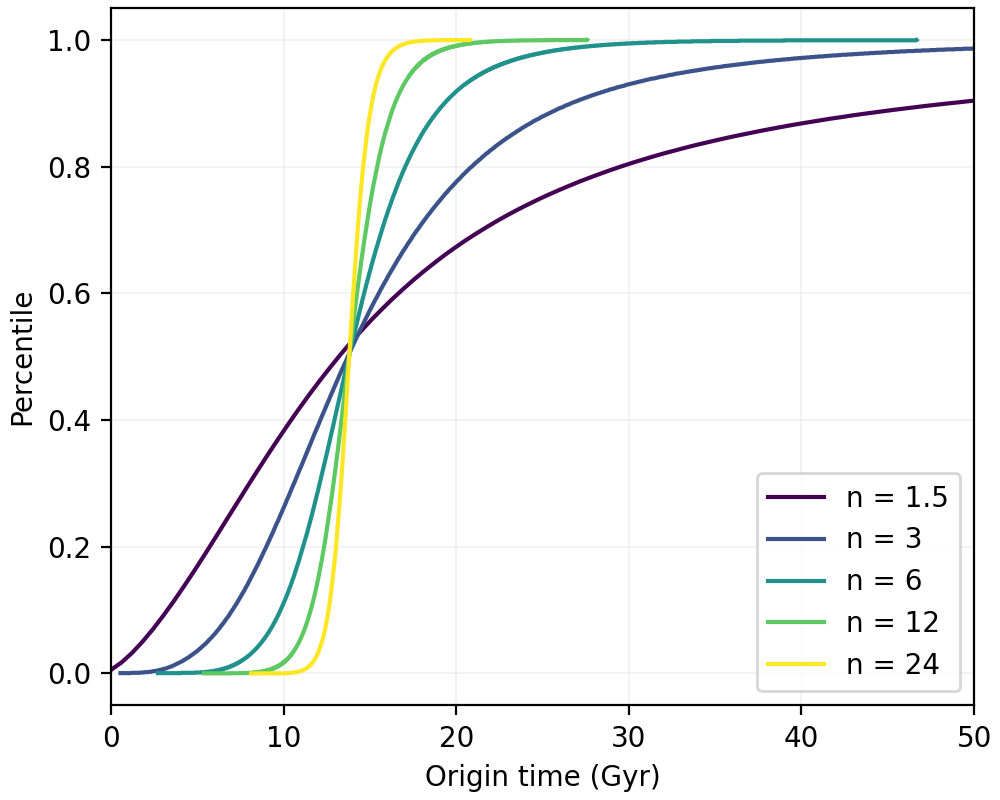}
\caption{C.d.f.s over GC origin clock-times, given uniform distribution on humanity’s rank. This is for speed $s=c$, where aliens are never seen.}
\label{fig:cdf-origin-times}
\end{minipage}\qquad\begin{minipage}{.45\textwidth}
\centering
\vspace{-15pt}\includegraphics[width=0.95\linewidth]{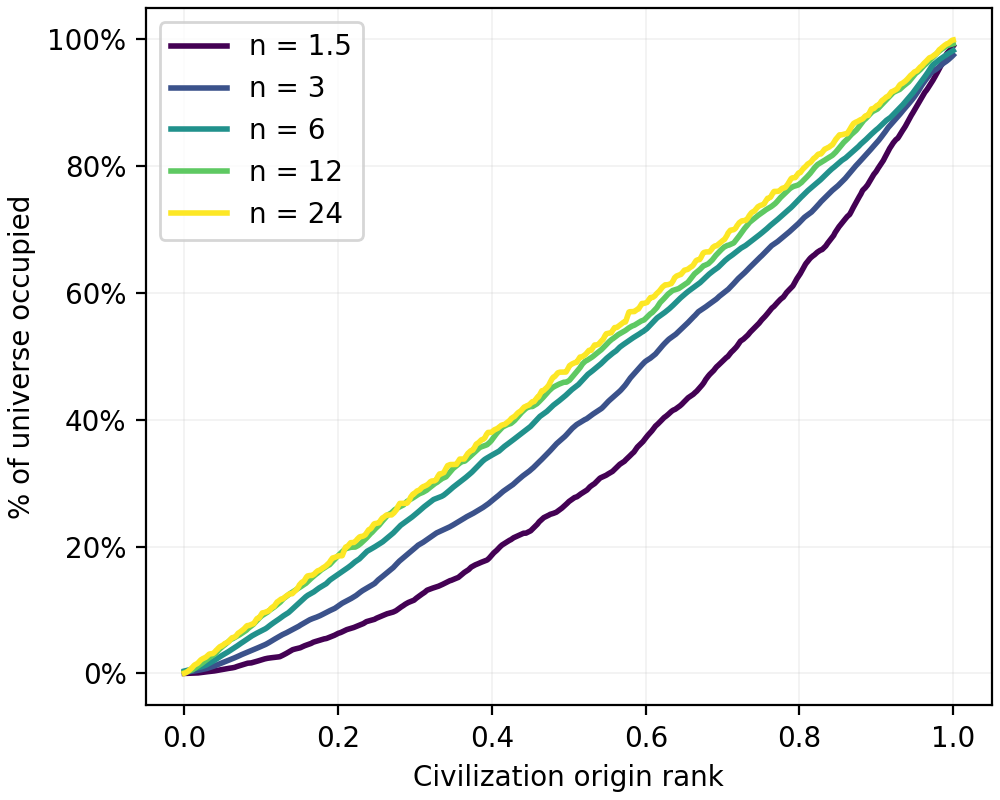}
\caption{Fraction of universe volume controlled by GCs, as a function of rank of GC origin time. }
\label{fig:volume-fractions}
\end{minipage}
\end{figure}

We will also see in Figure \ref{fig:volume-fractions} that, averaging over GC origin dates, a third to a half of the volume of the universe is controlled by GCs. So from a random location at such dates, one is likely to see large alien-controlled volumes. However, if the GC expansion speed is a high enough fraction of the speed of light, a selection effect, illustrated in Figure \ref{fig:illustration-selection}, makes it unlikely for a random GC to see such an alien volume at its origin date. If they were where we could see them, then they would be here now instead of us.

\begin{figure} 
\centering
\includegraphics[width=4in]{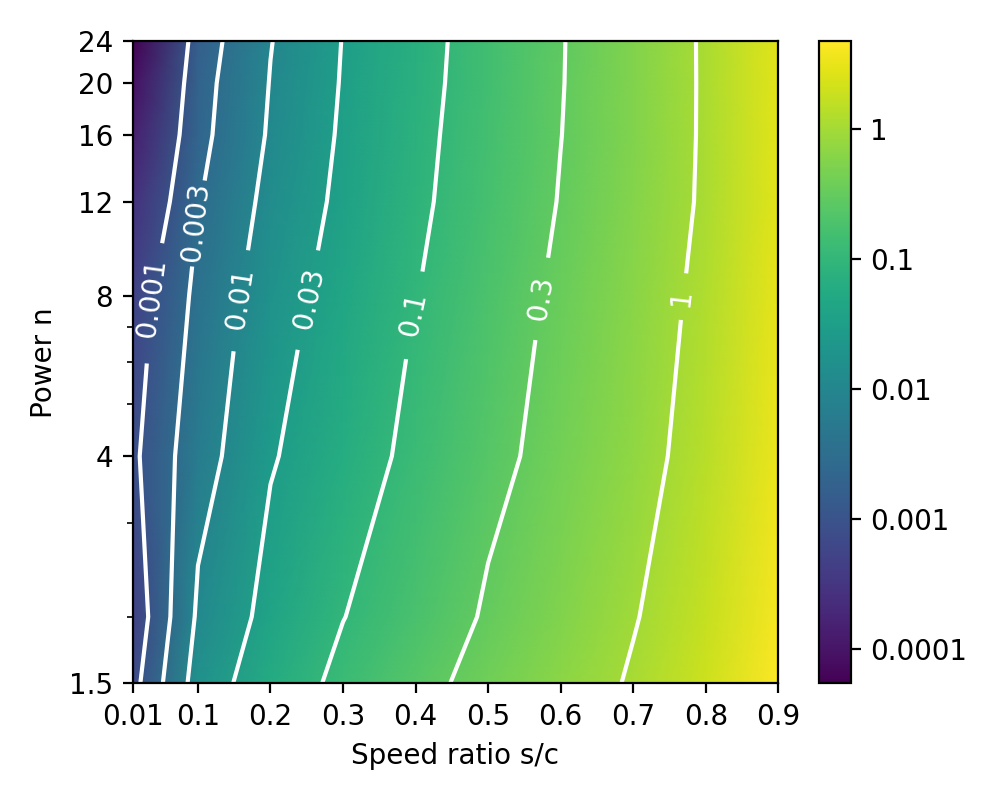}
\caption{Likelihood ratios, for volume-power and speed $(n, s/c)$ parameter pairs, regarding the observation that one sees no large alien-controlled volumes in the sky. To obtain a posterior distribution over these pairs, multiply this likelihood ratio by a prior for each pair, then renormalize.}
\label{fig:lik-ratios}
\end{figure}

Given prior estimates regarding the chances for various combinations of power $n$ and GC expansion speed ratio $s/c$, we can update such priors via the likelihood ratio of such a pair $(n,s/c)$ to predict our key evidence that we do not now see alien volumes. Figure \ref{fig:lik-ratios} show that likelihood ratio, i.e., the number of GC origin events that see no alien volumes, divided by the number that do. Speed ratios of ${s}/{c} < \unarysim {1}/{3}$ are greatly disfavored, especially for high powers. 

Updating on not seeing alien volumes should also update one toward estimating an earlier rank, as earlier GC are less likely to see alien volumes. Even so, this does not weaken the inference from Figure \ref{fig:lik-ratios} toward a high speed $s$.

This analysis, like most in our paper, assumes we would have by now noticed differences between volumes controlled or not by GCs. Another possibility, however, is that GCs make their volumes look only subtly different, a difference that we have not yet noticed. If even so we would have noticed being inside a GC-controlled volume, then our model still applies, except that we can't use current data to constrain the expansion speed $s$. 

\begin{figure}
\centering
\includegraphics[width=3.25in]{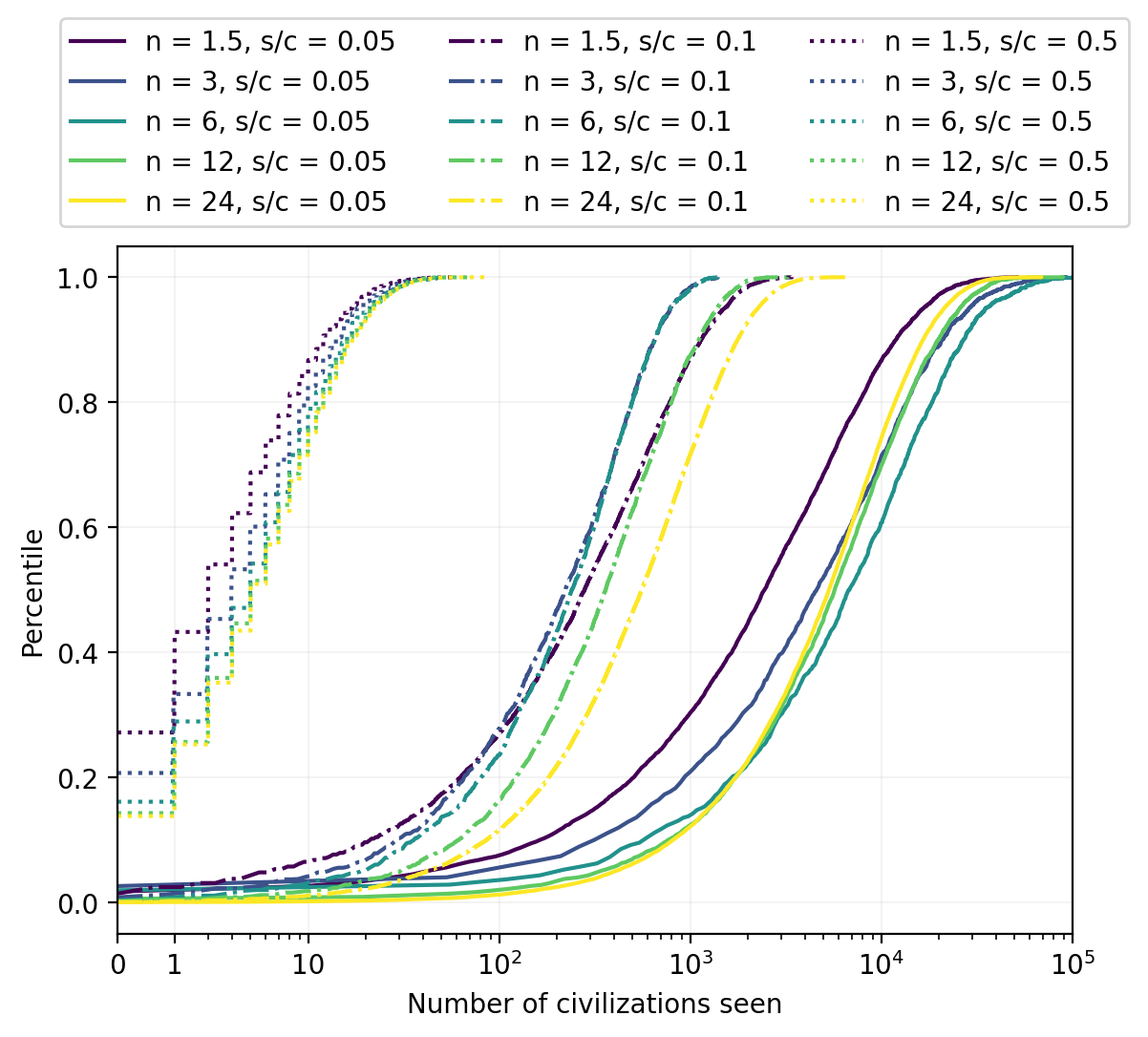}
\caption{Distributions over how many other GCs each one sees at its origin. At speed $s = c$, none see any others.}
\label{fig:cdf-num-seen}
\end{figure}

\begin{figure} 
\centering
\includegraphics[width=3.5in]{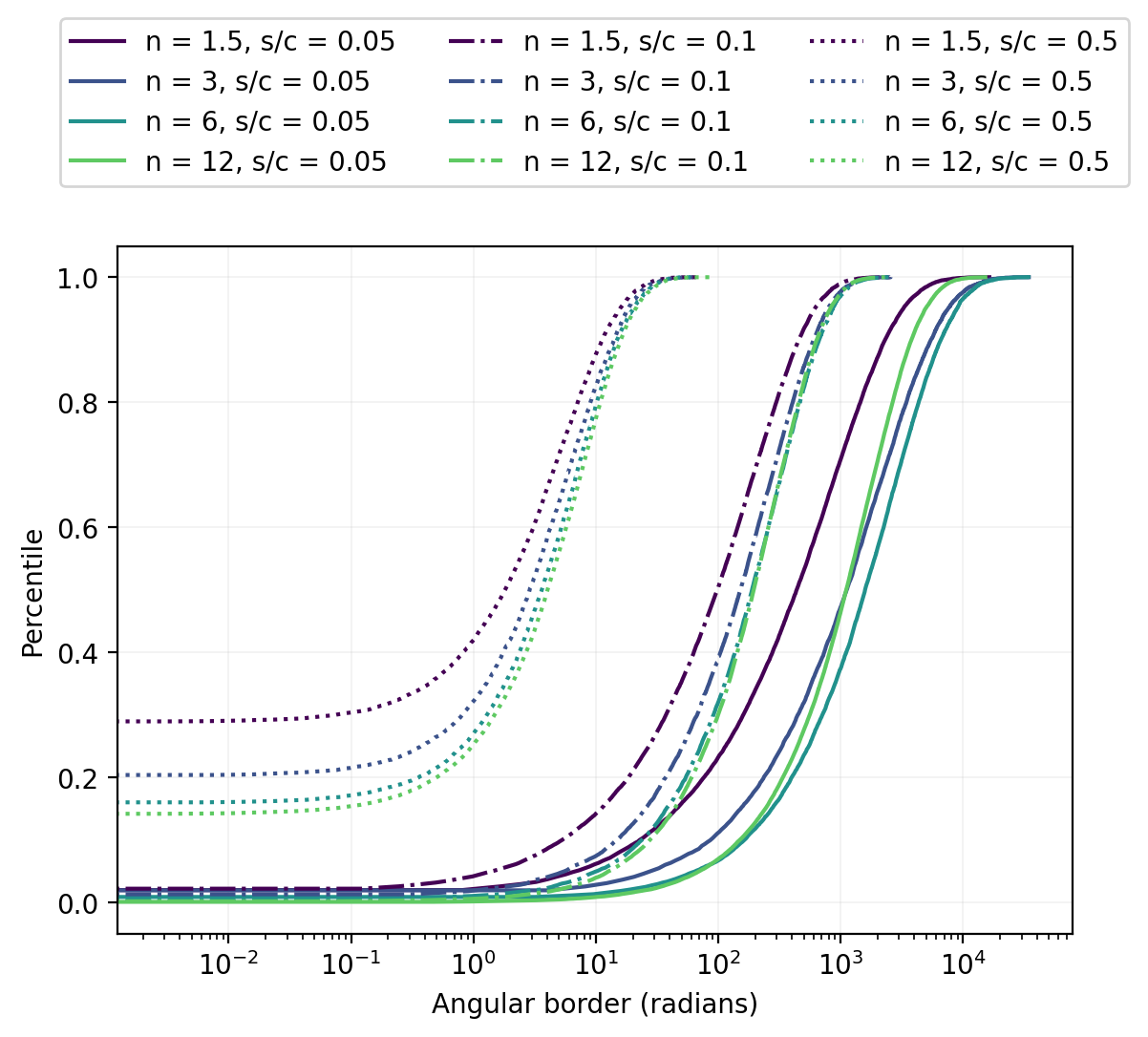}
\caption{The total angular length of GC borders on the sky at GC origin dates.}
\label{fig:border-lengths}
\end{figure}

In this case, there could be hope for astronomers to search the sky for subtle circular borders between GC volumes and surrounding volumes. Figures \ref{fig:cdf-num-seen} and \ref{fig:border-lengths} show how predicted distributions over the number of GC volumes and the total length of such sky borders vary with speed $s$ and power $n$.

\begin{figure} 
\centering
\centering
\includegraphics[width=2.6in]{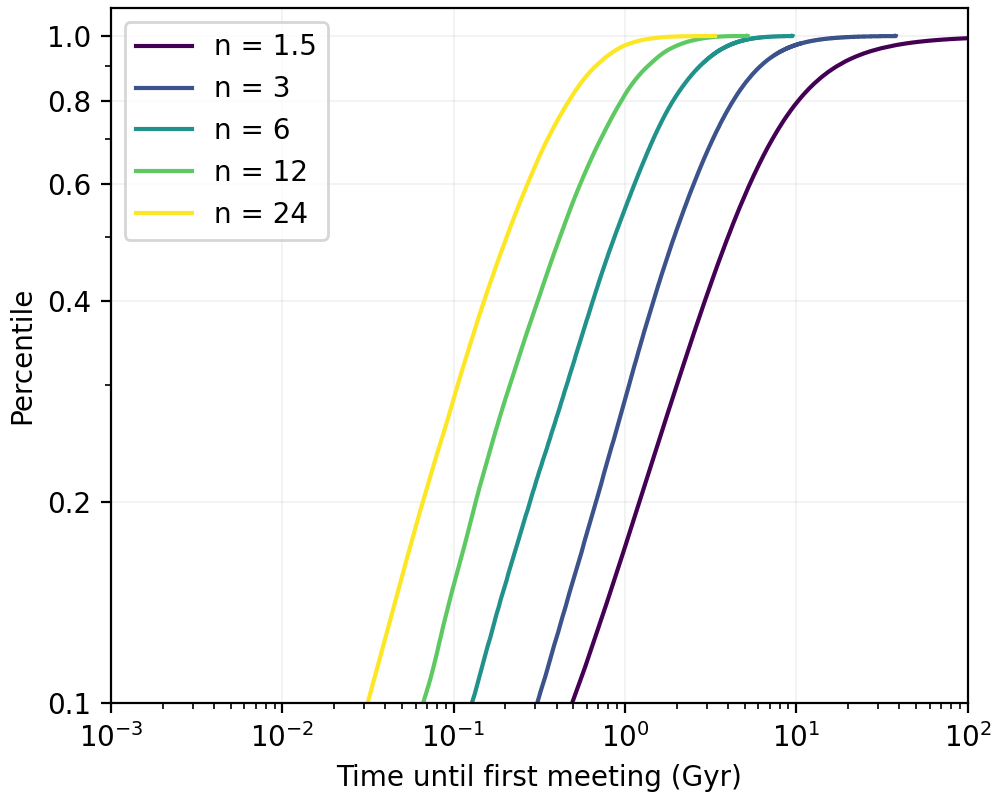}
\caption{C.d.f.s over clock-time until some of our descendants directly meet aliens.}
\label{fig:cdf-meet-times}
\end{figure}

\begin{figure}
\centering
\begin{minipage}{.45\textwidth}
\centering
\includegraphics[width=1.0\linewidth]{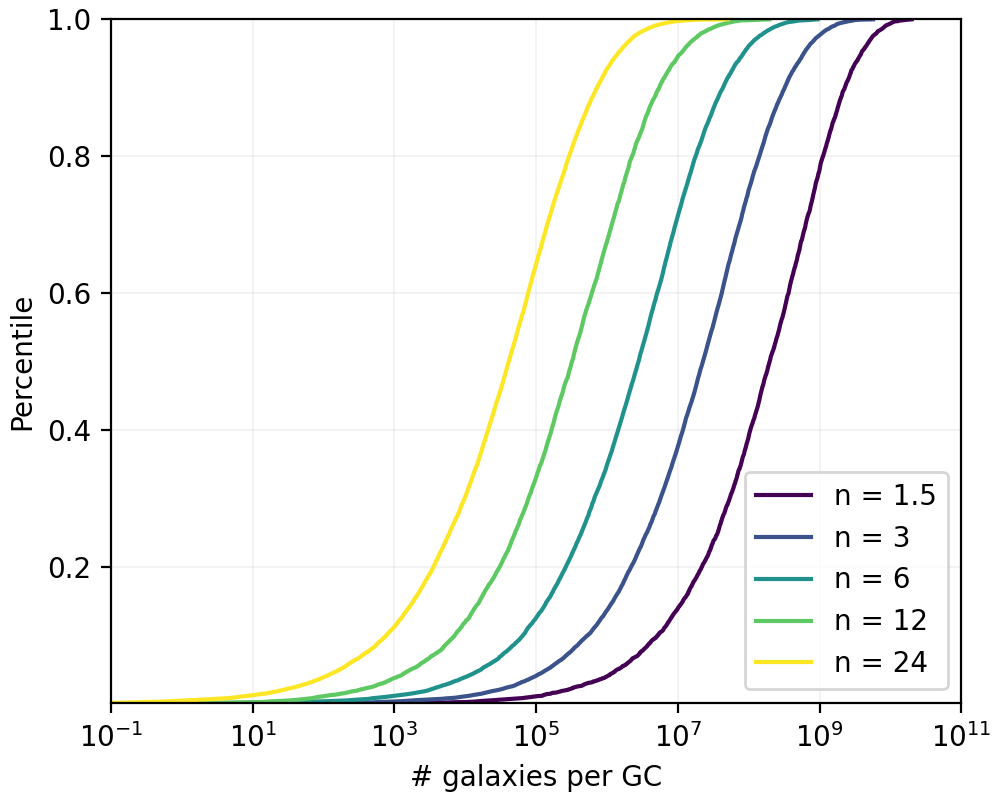}
\caption{Final galaxies per GC-controlled volume, if GCs retain initial volumes after meeting. This for $s=c$; others go as $(s/c)^3$.}
\label{fig:galaxies-per-civ}
\end{minipage}\qquad\quad\begin{minipage}{.45\textwidth}
\vspace{-10pt}\includegraphics[width=1.0\linewidth]{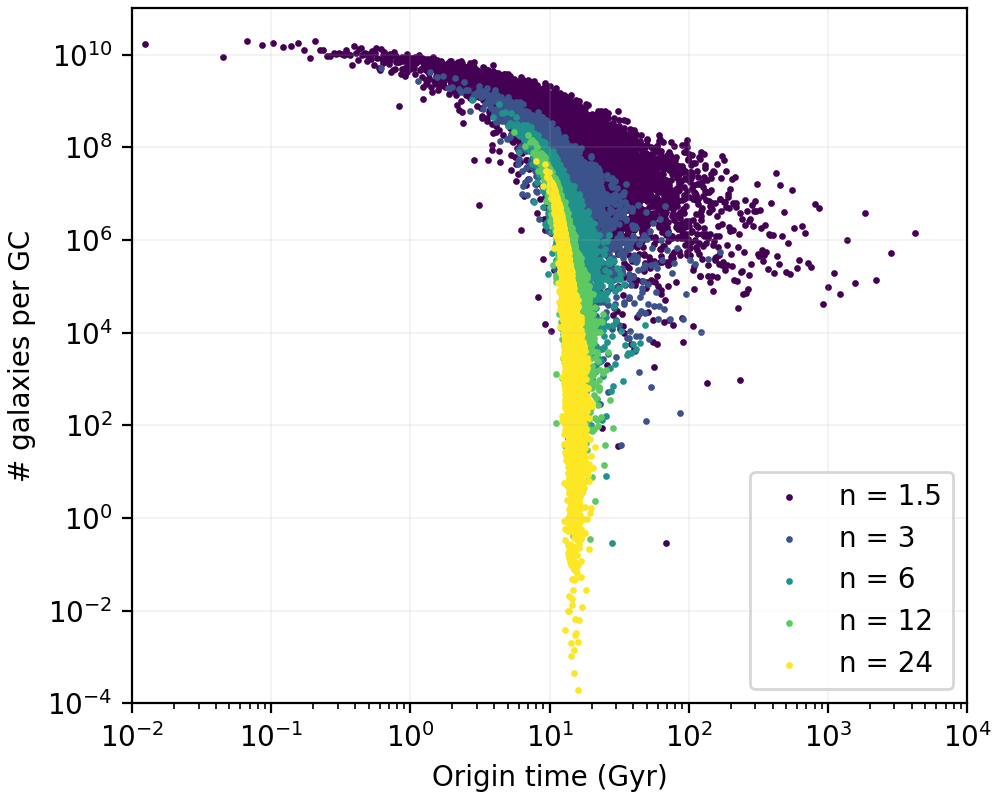}
\caption{Final galaxies per GC-controlled volume, versus the origin time of each GC. Results from one simulation run.}
\label{fig:origin-volume}
\end{minipage}
\end{figure}

\section{Simulation Results}
\label{sec:simresults}

We now show more graphs describing how distributions over GC statistics vary with volume-power $n$, and sometimes also with speed ratio $s/c$. Unless stated otherwise, these each come from averaging over five simulation repetitions, each with $W=s=c=1$ and $N=10^8$ sample GC origin events. Model-to-clock time mappings are made using only GCs who see no others at their origins, but all GCs are shown in the distributions, even GCs who do see others. Correctness of code has often been checked by comparing independent implementations. 

Figure \ref{fig:cdf-origin-times} shows clock GC origin dates, and regarding those origin events Figure \ref{fig:volume-fractions} shows GC-controlled volume fractions. Figure \ref{fig:cdf-meet-times} shows clock durations until a GC descendant meets aliens,  Figure \ref{fig:cdf-max-angle} shows the largest alien volume angle seen, and Figure \ref{fig:cdf-time-see} shows clock durations until a GC descendant sees aliens. 

Figure \ref{fig:galaxies-per-civ} shows distributions over the number of galaxies per GC at the simulation end, when GCs fill all volumes, under the assumption that GCs who meet stop and retain their volumes. We apparently live on a one-in-a-million-galaxies ``rare Earth'' \cite{ward2000rare}. 

Figure \ref{fig:origin-volume} shows how GC origin dates and volumes are related to each other. Note that the higher the volume-power, the more closely spaced are GC origin times, the fewer galaxies each GC encompasses, and the less time until our GC descendants would meet or see aliens. The earliest GCs tend to have the largest volumes.

Out of all the figures in this paper, only Figures \ref{fig:cdf-origin-times},\ref{fig:cdf-meet-times},\ref{fig:cdf-meet-times},\ref{fig:cdf-time-see},\ref{fig:origin-volume} depend on our assumption of a uniform distribution over humanity's origin rank among GCs.

\begin{table}[h]
    \centering
    \begin{tabular}{l|lll lll}
    & \multicolumn{3}{c}{\textit{Scenario:}} & \multicolumn{3}{c}{\textit{Scenario:}} \\
    & \multicolumn{3}{c}{$s/c=1/2, \  n=6$} & \multicolumn{3}{c}{$s/c=3/4, \ n=12$} \\
    \hline \vspace{-10pt}\\
    \textit{Percentile} & \textit{1\%} & \textit{25\%} & \textit{75\%} & \textit{1\%} & \textit{25\%} & \textit{75\%} \\
Origin (Gyr) & 8.99 & 15.25 & 21.24 & 10.20 & 13.46 & 16.02 \\
MinTillMeet (Gyr) & 0.019 & 0.488 & 2.226 & 0.006 & 0.188 & 0.882 \\
MinTillSee (Gyr) & 0 & 0 & 0 & 0 & 0 & 0.425 \\
MaxAngle & 0 & 0.132 & 0.908 & 0 & 0 & 0.313 \\
\% Empty & 0.010 & 0.320 & 0.830 & 0.010 & 0.290 &
0.810 \\
    \end{tabular}
    \caption{Specific numbers for two scenarios, $(n,s/c)$ = $(6, 1/2), (12, 3/4)$.}
    \label{tab:sim-numbers}
\end{table}

For those frustrated by difficulties in reading numbers off our many graphs, Table \ref{tab:sim-numbers} gives specific numbers for two scenarios.

\section{SETI Implications}
\label{sec:SETI}

\begin{figure}
\centering
\begin{minipage}{.45\textwidth}
\centering
\includegraphics[width=1.0\linewidth]{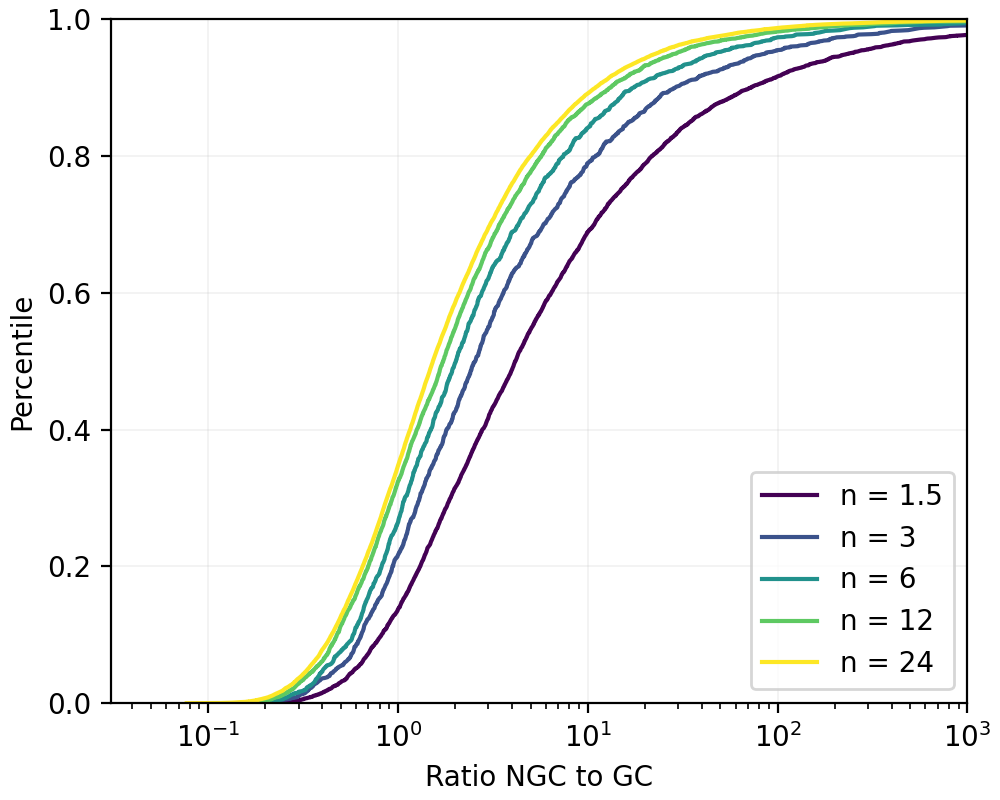}
\caption{Ratio $R$ required to expect one NGC ever anywhere within our past light cone. This is for $s=c$; others go as $(s/c)^3$.
}
\label{fig:ratio-r1}
\end{minipage}\qquad\quad\begin{minipage}{.45\textwidth}
\vspace{25pt}\includegraphics[width=1.0\linewidth]{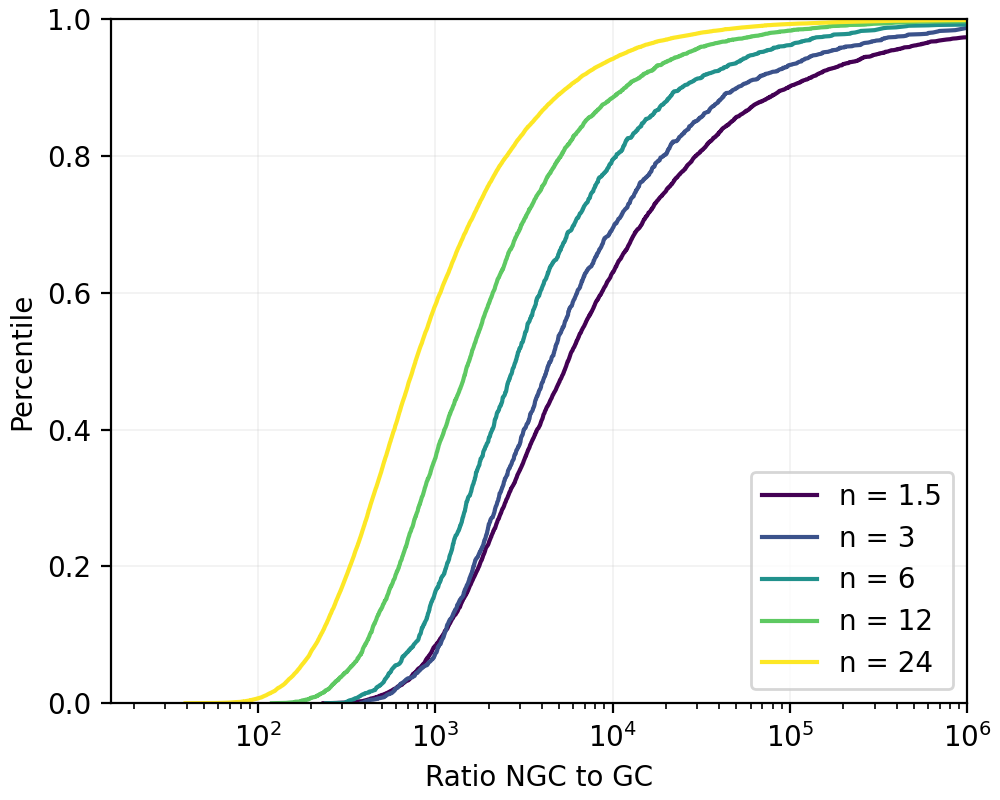}
\caption{Ratio $R$ required to expect one NGC intersecting with our past light cone. This is for $s=c$; others go as $(s/c)^2$. This is for an expected NGC lifetime $L = 1$Myr; others go as $L^{-1}$.}
\label{fig:ratio-r2}
\end{minipage}
\end{figure}

\begin{figure}
\centering
\begin{minipage}{.45\textwidth}
\centering
\includegraphics[width=1.0\linewidth]{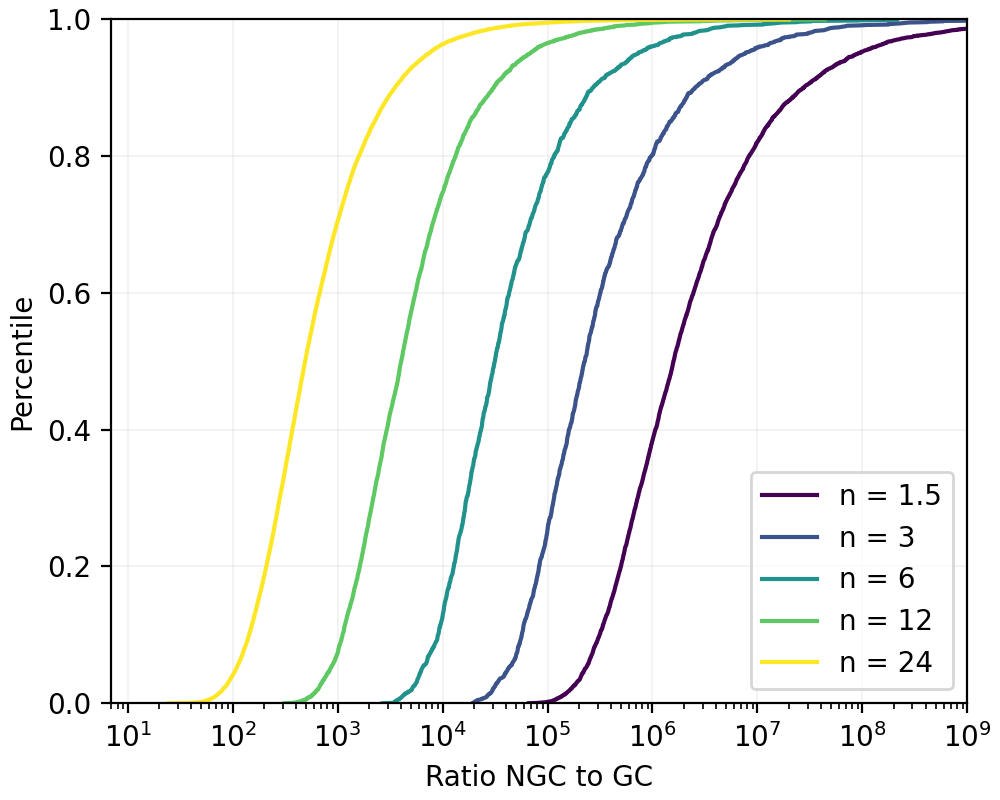}
\caption{Ratio $R$ required to expect one NGC has ever existed in our galaxy. That is, both within our past light cone, and also in our galaxy. This is for $s=c$; others go as $(s/c)^3$. }
\label{fig:ratio-r3}
\end{minipage}\qquad\quad\begin{minipage}{.45\textwidth}
\vspace{1pt}\includegraphics[width=1.0\linewidth]{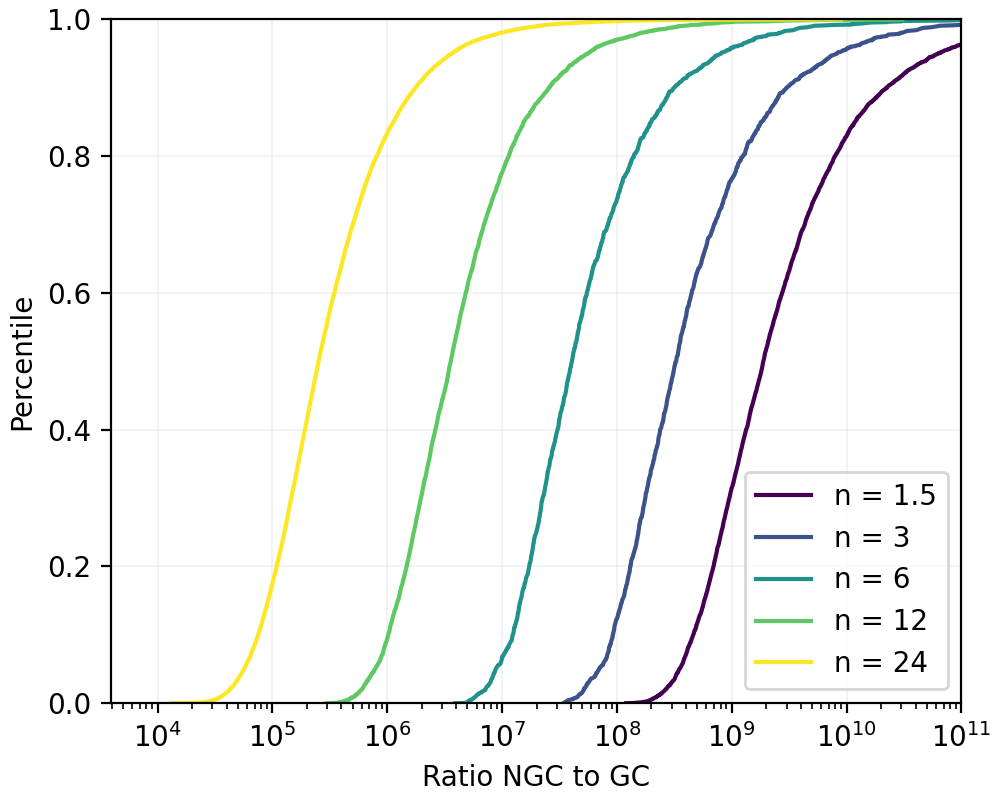}
\caption{Ratio $R$ to expect one NGC ``now'' in our galaxy. That is, intersects our past light cone, and in our galaxy. This is for $s=c$; others as $(s/c)^2$. This is for lifetime $L = 1$Myr; others go as $L^{-1}$.}
\label{fig:ratio-r4}
\end{minipage}
\end{figure}

The results we have shown so far have been on grabby civilizations (GCs), including where and when they exist, and how we might see or meet them. Let us now consider non-grabby civilizations (NGCs), who we have said are $R$ times more common than grabby ones; each NGC has a $1/R$ chance to soon birth a GC. This unknown ratio $R$ sets the chances for \textit{both} humanity's grabbiness in the next ten million years, and \textit{also} for success in SETI (search for extraterrestrial intelligence) efforts. 

On human futures, we can think of humans today as a NGC, or soon to become one. While we might not want to become a GC, many of the scenarios in which we do not are because we can not. For example, we may go extinct, or become permanently and strongly limited. So we might at least like to have the option to become grabby. Thus human future optimists tend to think that our descendants have a decent chance to birth a GC, and thus have low estimates of $R$. (One author's Twitter poll gave a median for humanity of $R\unarysim2.5$.)

In contrast, high estimates of $R$ are needed to expect SETI success anytime soon. The four Figures \ref{fig:ratio-r1},  \ref{fig:ratio-r2},  \ref{fig:ratio-r3}, and \ref{fig:ratio-r4} show distributions over the ratio $R$ required to expect one active NGC in these four different spacetime regions where SETI might look: 

\vbox{%
\begin{itemize}
\item Within our past lightcone (Figure \ref{fig:ratio-r1})
\item Intersecting our past lightcone (Figure \ref{fig:ratio-r2})
\item Before ``now'' in our galaxy (Figure \ref{fig:ratio-r3})
\item ``Now'' in our galaxy (Figure \ref{fig:ratio-r4}) 
\end{itemize}}

NGCs in regions further down this list seem easier for us to see if they are there, but require larger ratios $R$ for any to be there to see. For example, human future optimists who see $R > 10^3$ as implausible should agree with SETI pessmimists who say there have probably never been aliens in our galaxy \cite{hart1975explanation}.

To expect ten times as many NGCs in a region, you need a ratio $R$ ten times higher. GCs lower in the percentile rank $r$ have a smaller required $R_r$, and would for a true ratio $R$ expect $R/R_r$ NGC per region.
Figures \ref{fig:ratio-r1}, \ref{fig:ratio-r2},  \ref{fig:ratio-r3}, and \ref{fig:ratio-r4} show results for $s=c$; for $s=c/2$, ratios 8 or 4 times smaller are required. Figures \ref{fig:ratio-r2} and \ref{fig:ratio-r4} are for expected NGC lifetimes of $L = 1$Myr; for a lifetime ten times smaller, you need a ratio ten times larger. Thus the longevity of alien technosignatures is a key to SETI success \cite{balbi2021longevity}. For example, aliens who build beacons that function long after their civilizations die may be much easier to see \cite{benford2010messaging,lacki2020lens}. 

Note that Figures \ref{fig:ratio-r1} and \ref{fig:ratio-r3} are actually versions of Figures \ref{fig:ratio-r2} and \ref{fig:ratio-r4} in the limit of indefinite lifetime $L$. Note also that higher powers $n$ usually require lower ratios $R$, and thus offer more room for SETI optimism, though high $n$ also strengthens the human earliness puzzle, and so pushes more for the grabby aliens model.

At our middle estimate volume power of $n=6$, a NGC to GC ratio of over one thousand is required to expect even one NGC active on our past light cone, and a ratio of over ten thousand is required to expect one anywhere in the past of our galaxy. Furthermore, a ratio of over ten million is required to expect one NGC active now in our galaxy. (And that’s for a million year NGC lifetime; shorter lived NGCs require even higher ratios). Yet today SETI struggles to see NGC techno-signatures for even a tiny fraction of the stars in our galaxy. Overall these seem discouraging results for SETI.

Note that the widely-used ``self indication'' prior for indexical probability analysis favors larger values of $R$ in proportion to $R$, at least relative to the chances that one might estimate neglecting anthropic considerations \cite{grace2010sia, olson2021implications}.

Note also that GC-controlled volumes, like other volumes, are mostly empty and probably transparent, and that we may have not yet learned how to see their differences; Section \ref{sec:expansionspeed} shows the angle lengths of the circular borders we might then find in our sky.

\section{Conclusion}
\label{sec:conclusion}

A literature has modeled the evolution of life on Earth as a sequence of ``hard steps'', and has compared specific predictions of this model to Earth’s historical record. This model seems to roughly fit, and supports inferences about the number of hard steps so far experienced on Earth. While some argue that other processes besides hard steps also happened (like the easy, delay, try-once, and multi-try steps discussed in Section \ref{sec:hardsteps}), we know of no published cricitism of the basic idea that some hard steps occured. That is, this hard steps model seems widely accepted.

Yet we suggest that this standard hard-steps model does not seem to have been taken sufficiently seriously. For example, the literature that estimates the timing of the appearance of advanced life has only once included the key hard-steps timing power law. In Section \ref{sec:appearance} we have presented a simple model which includes this effect, and in Section \ref{sec:humansearly} we find that according to this model humanity today seems to be quite early, unless one assumes both a rather low planet-power \textit{and} a very restrictive limit on habitable planet lifetimes. 

Related literatures have also apparently not considered applying this hard-steps-based power law to larger volumes like galaxies, instead of to just planets. Sterilizing explosions reduced galactic habitability eons ago, and some have suggested that this supports a scenario wherein advanced civilizations are far more common today than they were eons ago \cite{cirkovic2008astrobiological}. But such authors do not seem to have realized that, with a sufficiently high power, a volume-based hard steps power law directly produces a similar effect.

Nor do most prior authors seem to have noticed that a scenario wherein advanced civilizations grab most of the available volumes can robustly explain humanity's early arrival. While others have offered explanations based on the assumption that the evolution advanced life must follow a path close to Earth's path, our deadline explanation allows for a very wide range of evolutionary paths and contexts. 

To formalize this argument, we have presented in this paper a very simple model (plausibly over-simplified in fact) of what we call ``grabby'' civilizations (GC), a model that is a modest variation on models by S. Jay Olson, who recently pursued a similar approach, minus the hard steps power law \cite{olson2015homogeneous,olson2016visible,olson2017estimates,olson2018expanding,olson2018long,olson2020likelihood}. 

In our model, GCs are born according to a volume-based power law, and once born they simply expand at a constant speed relative to local materials. We show that this power law is often at least a crude approximation to a more realistic model. This expansion speed and the two parameters of this power law are the only three parameters of our model, each of which can be estimated from data to within roughly a factor of four.

The hard-steps in Earth history literature helps to estimate the power, and our current date helps to estimate the power law timescale. Furthermore, the fact that we do not now see large alien-controlled volumes in our sky, even though they should control much of the universe volume now, gives us our last estimate, that aliens expand at over half of lightspeed. Given estimates of all three parameters, we have in this paper shown many model predictions regarding alien timing, spacing, appearance, and the durations until we see or meet them. And we have shown how optimism regarding humanity's future is in conflict with optimism regarding SETI efforts.

Being especially simple, our model is unlikely to be an exact representation of reality. So future research might explore more realistic variations. For example, one might more precisely account for the recent exponential expansion of the universe, and use a more accurate appearance function instead of our power law approximation. Instead of being uniform across space, the GCs birth rate might be higher within galaxies, higher within larger galaxies, and follow their typical spatial distributions. A GC expansion might take a duration to bring its full effect to any one location, and the GC expansion speed might vary and depend on local geographies of resources and obstacles. Finally, GC subvolumes might sometimes stop expanding or die, either spontaneously or in response to local disturbances.

Note, however, that in many cases we may not have data available to estimate the extra parameters that these extended models would introduce. A virtue of our over-simplified model is that we can estimate all its parameters using available data, and show the full range of variation of its behavior.


\clearpage


\bibliographystyle{Science}


\section*{Acknowledgments}

For support, we thank Steve Kuhn, Center for the Study of Public Choice at George Mason University, and Salem Center for Policy at University of Texas McCombs School of Business.

For comments and discussion, we thank: Scott Aaronson, David Brin, Andrew Hanson, Chris Hibbert, Anders Sandberg, Carl Shulman, David Deutsch, Kristian Moffat, Adam Brown, Stuart Armstrong, S. Jay Olson, Sara Walker, Paul Davies. and participants in the February 8, 2021 online meeting of the Foresight Institute's Intelligent Cooperation Group, and participants in the March 5, 2021 online seminar at the Future of Humanity Institute, Oxford University. 


\clearpage

\section{Appendix}

\subsection{Power Law Test Appendix}
\label{sec:testpowerlaw}

Our grabby aliens model makes the conveniently-simple assumption that advanced civilizations arrive according to a volume-based power law in time. But while we have seen how such a power law can apply during an individual planet lifetime, and while we have included a planet-based power law in our more general volume-based expression $\alpha(t)$ of Equation (\ref{eq:alpha-t-cdf}), that formula is not itself a power law. So how well does a power law approximate this appearance function? 

As Equation (\ref{eq:alpha-t-cdf}) is the time integral of a product of several powers laws and an exponential decay, we expect an overall power law to fit better fit at early times, before the exponential decay gets strong. We also expect a better fit when contributing powers are larger, such as for larger planet-powers $n$, and when powers can act over longer time periods, such as for larger maximum planet lifetimes $\bar{L}$. 

\begin{figure}[h]
    \centering
    \includegraphics[width=.98\linewidth]{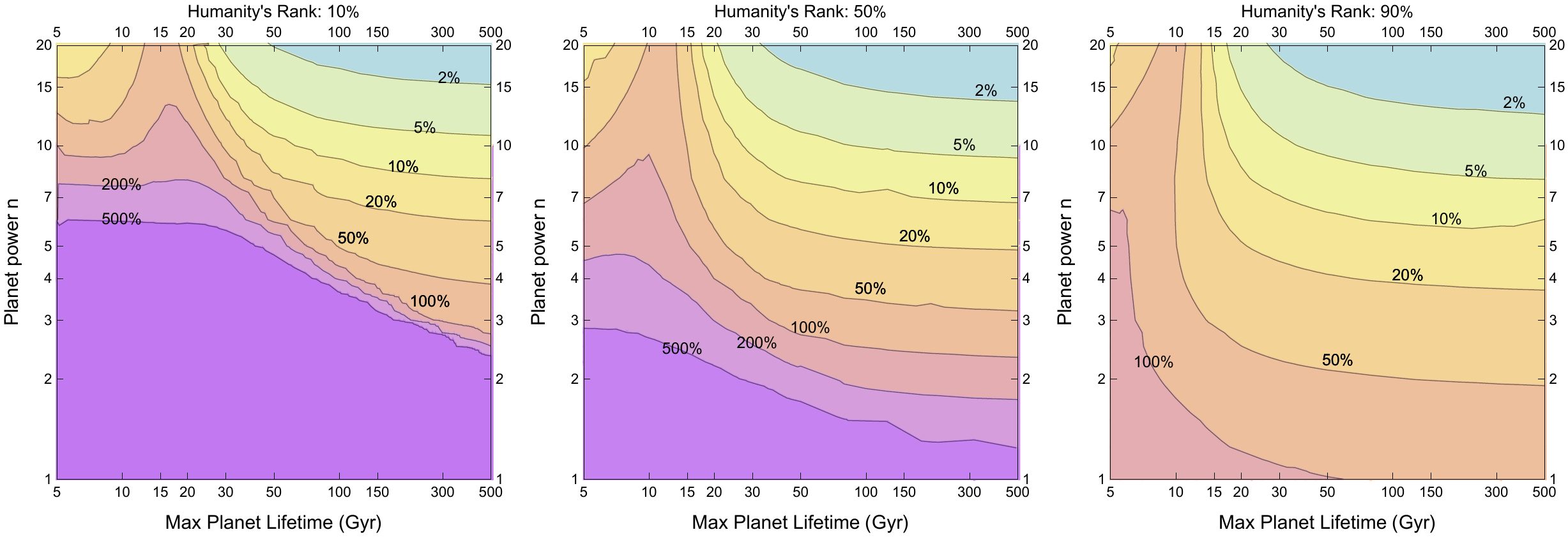}
    \caption{Average absolute error between natural logarithms of Equation \ref{eq:alpha-t-cdf}'s appearance function $\alpha(t)$ and a best approximating power law $(t/k)^{v-1}$, for volume-power $v$, averaged over GC origin dates from a $(t/k)^{v-1}$ simulation with humanity at percentile rank $r$. This percent error varies with planet-power $n$, max lifetime $\bar{L}$, and rank $r$. A 1\% absolute error is where the ratio of $\alpha(t)$ to $(t/k)^{v-1}$ is $1.01$ or $0.99$, while a 100\% error is where that ratio is $2$ or $0.5$.}
    \label{fig:model-error}
\end{figure}

\begin{figure}[h]
    \centering
    \includegraphics[width=.98\linewidth]{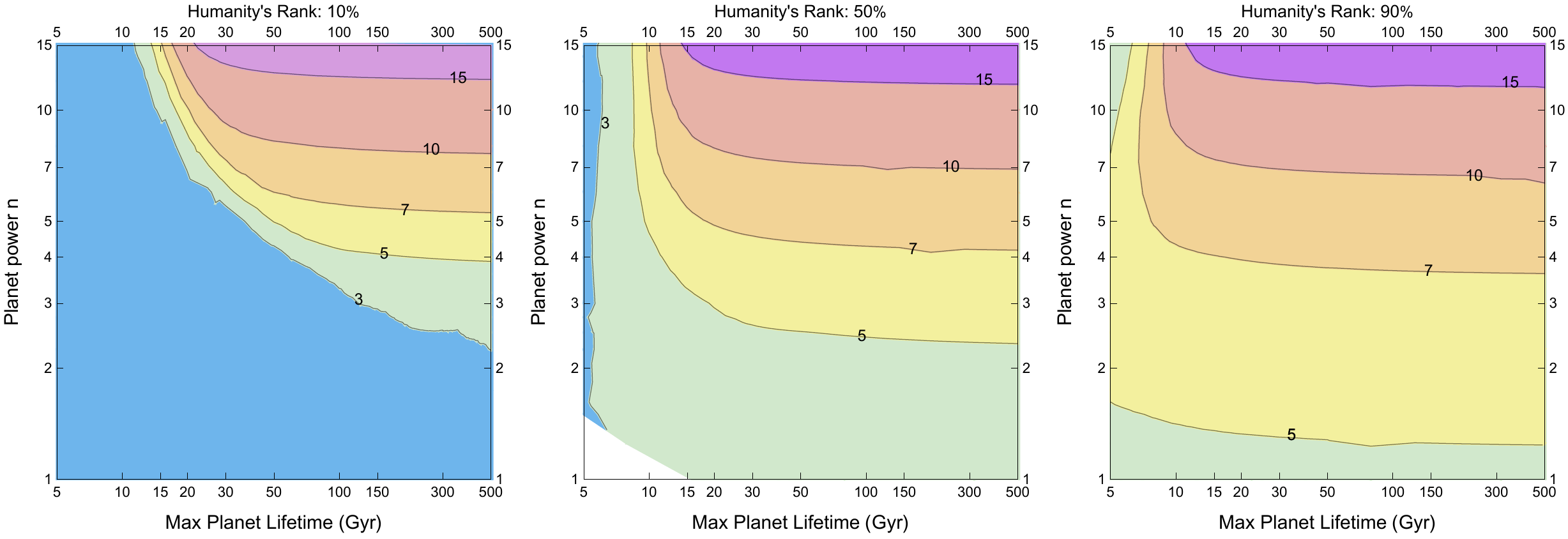}
    \caption{Best-fit volume-power $v$, found varying the same parameters and via the same approximation process as Figure \ref{fig:model-error}.}
    \label{fig:model-power}
\end{figure}

To study this error, we have approximated many particular arrival chance functions $\alpha(t)$ by the power law $(t/k)^{v-1}$, where $v$ is the volume-based c.d.f. power. Specifically, given a particular $\alpha(t)$ set by particular parameter values, including a planet power $n$, we vary the power law parameters $v,k$ to minimize, over GC origin dates $t_i$, the average absolute percent error $\ln((t_i/k)^{v-1}/ \alpha(t_i))$. These origin times $t_i$ are clock times taken from a $(t/k)^{v-1}$ grabby aliens simulation using that same clock time power $v$, and assuming that humanity is at rank $r$ among its GC origins. 

Regarding this power-law approximation, Figure \ref{fig:model-error} shows this percent error, and 
Figure \ref{fig:model-power} shows the best fit volume power $v$. Both figures vary planet-power $n$, max lifetime $\bar{L}$, and humanity rank $r$. Note that as we could also have varied the SFR parameters $\varphi$ and $\chi$, these graphs are only representative of more possibilities. 

As expected, the error is smaller when planet-power $n$ is larger, max lifetime $\bar{L}$ is larger, and when humanity's rank $r$ is higher, as in this case most GCs appear at much earlier dates than today, and thus before the 12Gyr peak we've assumed for the habitable star formation rate function $\varrho(t)$. 

At high humanity ranks $r$, a power law is almost always a decent approximation, and the volume power is larger than the planet power. At middle ranks $r$, the approximation is at least crude (i.e., on average within a factor of two) for most of this 2D parameter space. Such as for planet-powers of $n \geq 3$ and lifetimes $\bar{L} > 50$Gyr, or when planet-power $n \geq 10$. Volume powers remain higher than planet powers for all but the shortest lifetimes. 

A power law is a worse approximation to GC appearance, however, at low humanity ranks $r$, unless the planet powers is very large ($n \geq 10$), or unless both the power $n$ is substantial and planet lifetime $\bar{L}$ is long. In this case, most GCs appear after the habitable SFR peak of 12Gyr, and so appear in the SFR exponential decay region. If our analysis is less reliable when humanity is one of the earliest GCs to arise, these parts of our graphs become less trusthworthy: early origin ranks, later origin dates, longer times til see or meet aliens, and more galaxies per GC.

Thus we see that for a large fraction of the relevant parameter space, a general $t^n$ power law is an at least a crude appoximation of the great filter, not just regarding the appearance of advanced life on individual planets, but also regarding (limited-enough periods of) much larger co-moving volumes that contain changing mixes of planets and other possible oases. 

\subsection{Miscellaneous Appendix}
\label{sec:miscappendix}

\begin{figure}[h]
    \centering
    \includegraphics[width=3.5in]{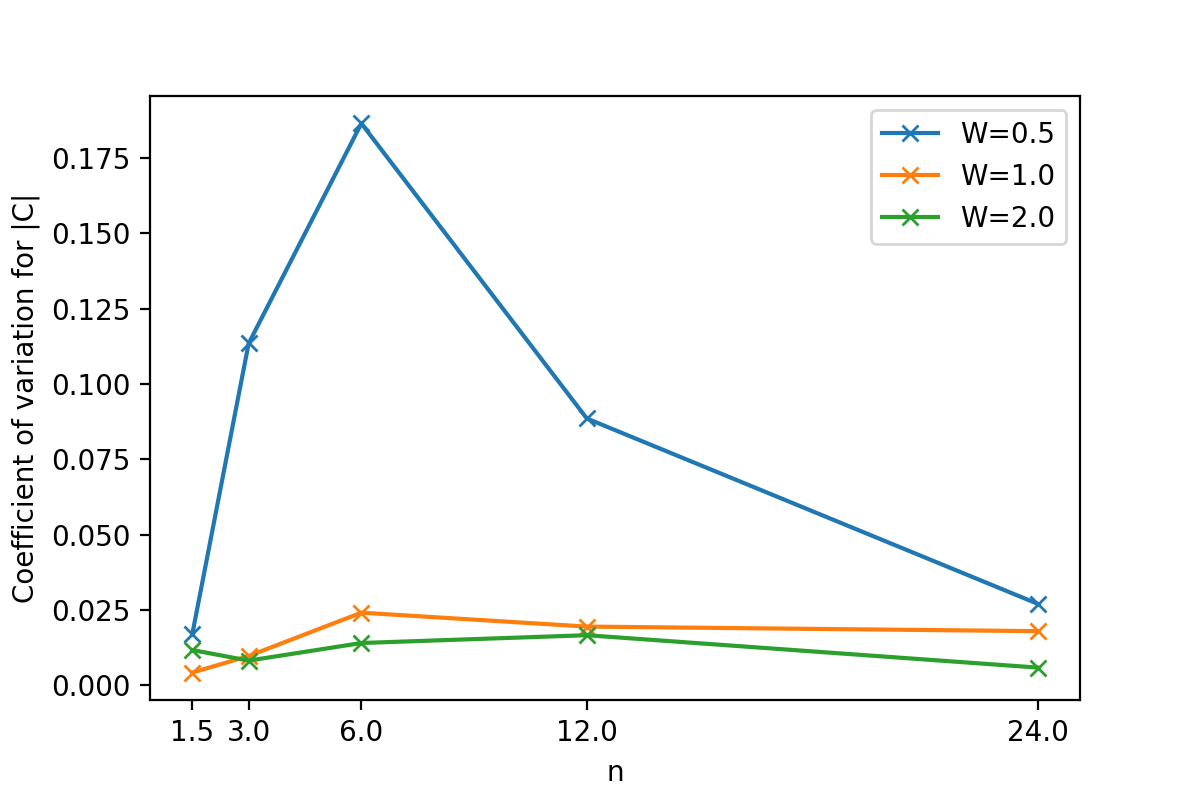}
    \caption{Testing if W=1 is sufficiently large, when s=1.}
    \label{fig:test-L1}
\end{figure}

One might worry about simulation border effects due to using speed $s=1$, a unit time interval $[0,1]$, and a wrap-around metric in a unit width box $[0,1]^D$. Figure \ref{fig:test-L1} varies box width by using a box $[0,W]^3$. It seems that width $W=1$ is sufficiently large.

\begin{figure} 
    \centering
    \includegraphics[width=3in]{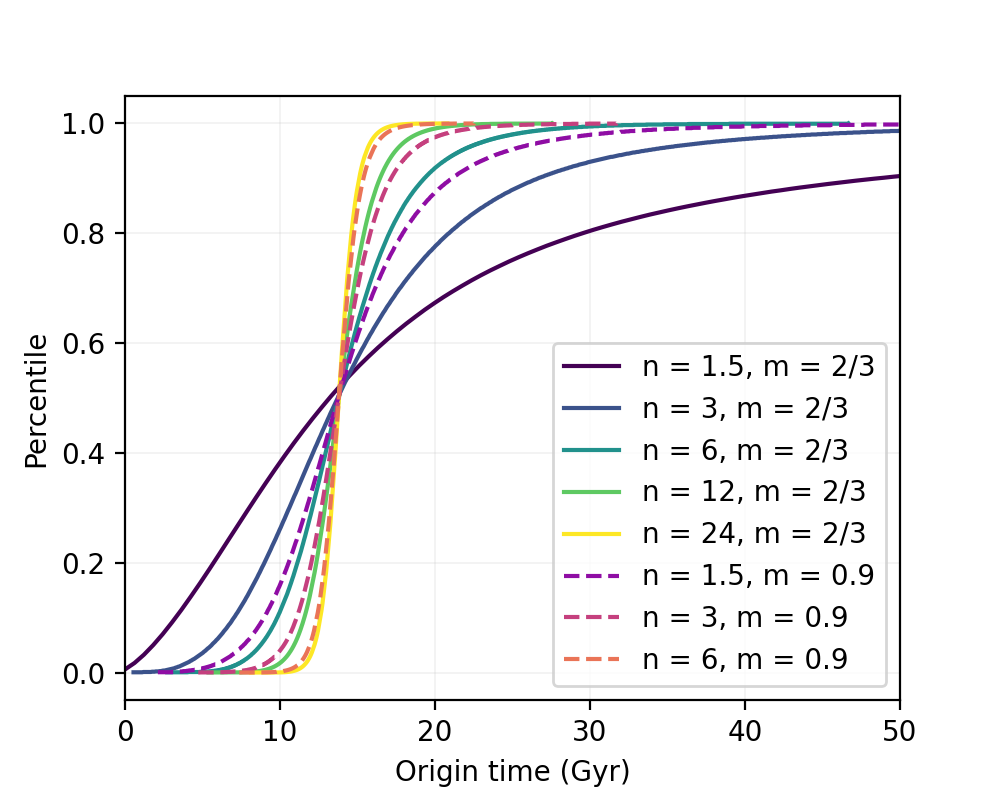} \\
    \includegraphics[width=3in]{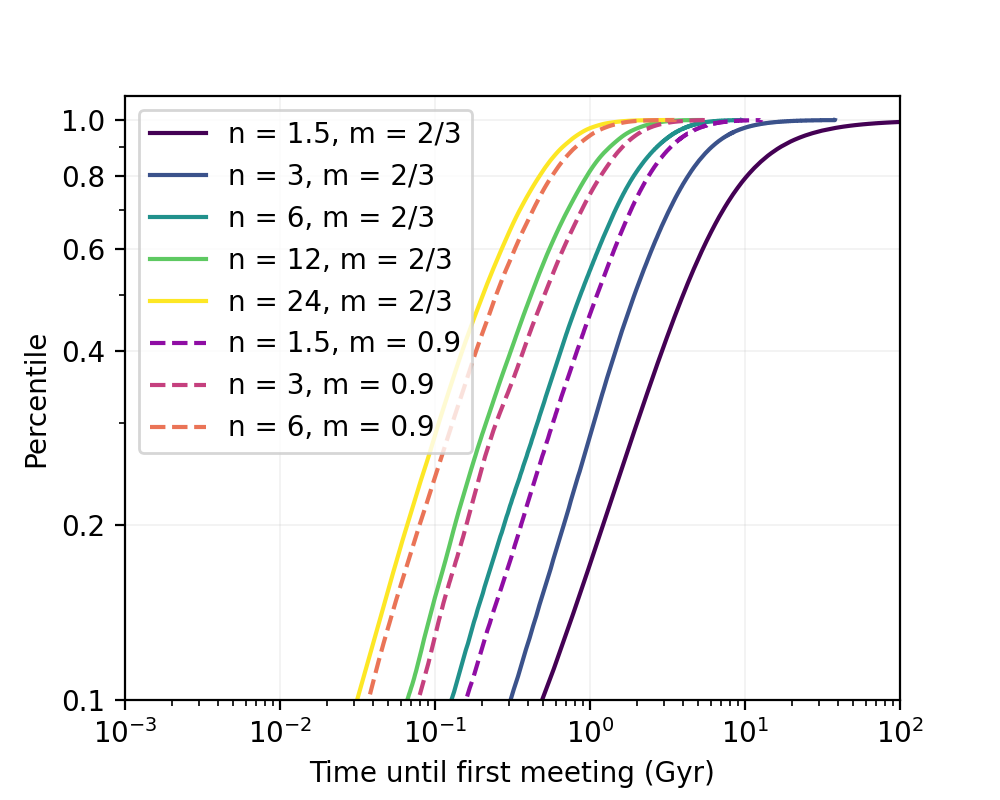} \\
    \hspace{-40pt}\includegraphics[width=3in]{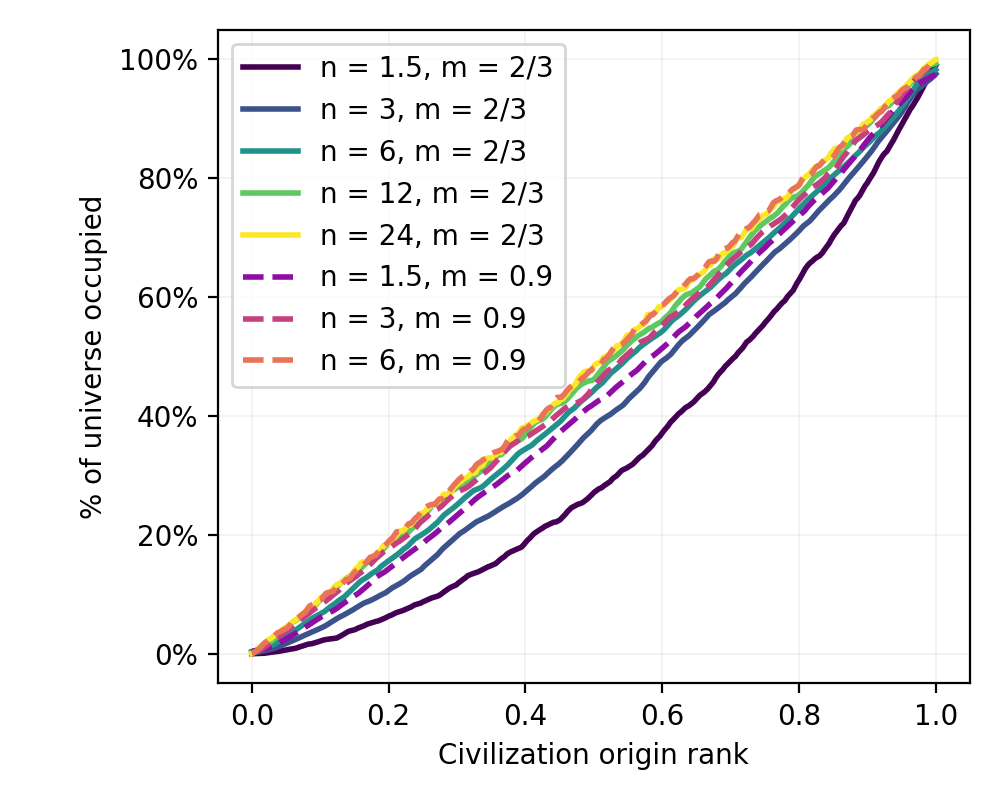}
    \caption{Comparing results for cosmological scale factor powers of $m = 2/3$ and $m = 0.90$.}
    \label{fig:different-m}
\end{figure}

\begin{figure} 
    \centering
    \includegraphics{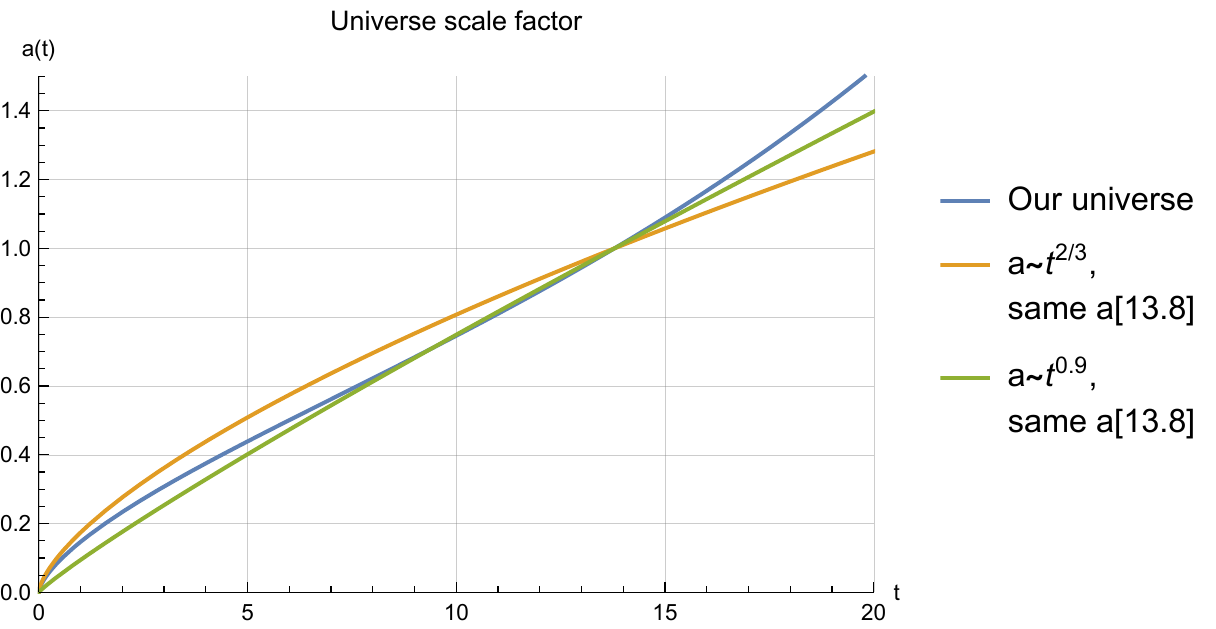}
    \caption{Cosmological scale factor over time in Gyr, in reality and as assumed in this paper.}
    \label{fig:scale-factor}
\end{figure}

As discussed in Section \ref{sec:cosmology}, most of our simulations have assumed a power law cosmological scale factor of $a(t) = t^m$, with $m = 2/3$. Figure \ref{fig:different-m} shows how some of our results change when $m = 0.9$ instead. It seems that the main effect is to, in effect, increase the clock-time volume-power $n$, as we expect from the relation $\eta = n/(1-m)$. Figure \ref{fig:scale-factor} shows how both of these power law approximations compare to the actual scale factor $a(t)$. 

\begin{figure} 
    \centering
    \includegraphics[width=6in]{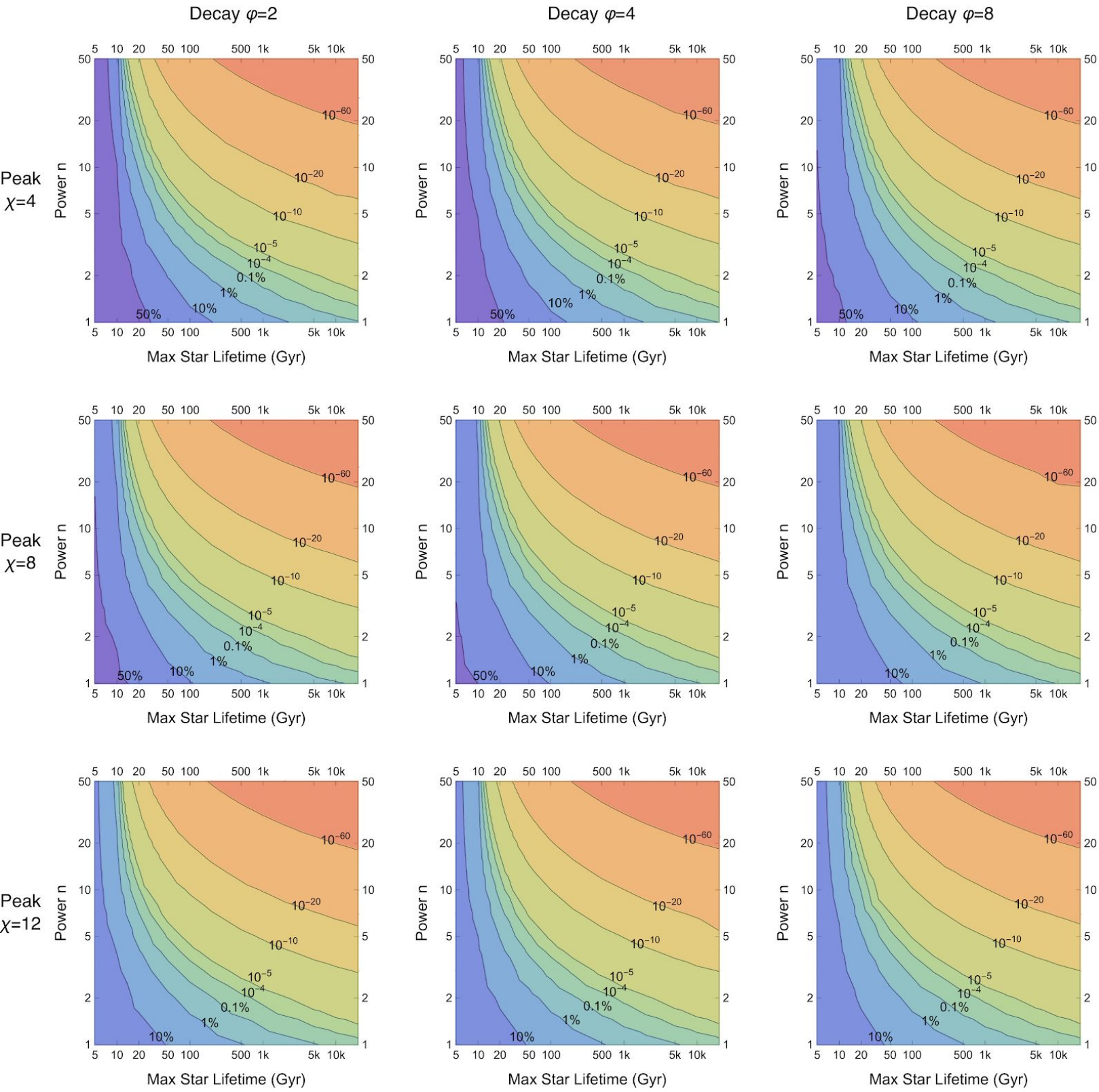}
    \caption{Percentile rank of today’s $13.8$ Gyr date within the distribution of advanced life arrival dates, as given by equation (1), assuming MFP $\kappa = 0$. Nine diagrams show different combinations of habitable SFR peak $\chi$ and decay $\varphi$, while each diagram varies planet-power $n$ and max habitable planet lifetime $\bar{L}$.}
    \label{fig:robustness-earliness}
\end{figure}

Figure \ref{fig:robustness-earliness} tests the robustness of our earliness estimates to varying the habitable SFR peak, by comparing the three values of peak $\chi$ in $4,8,12$ Gyr, all given MFP $\kappa =0$. 

\end{document}